\pdfoutput=1 
\documentclass[final,1p,times]{elsarticle}

\usepackage{graphicx}

\usepackage{amsmath,amssymb}

\usepackage[english]{babel}

\usepackage{siunitx}
\def\atm{atm}
\def\byte{B}

\usepackage[retainorgcmds]{IEEEtrantools}

\biboptions{comma, square, numbers, sort&compress}
\usepackage[linkcolor=blue, citecolor=blue, colorlinks=true, urlcolor=blue]{hyperref}
\usepackage{hypernat}

\hypersetup{
pdfauthor={Kyle E Niemeyer and Chih-Jen Sung},
pdftitle={Accelerating moderately stiff chemical kinetics in reactive-flow simulations using GPUs},
pdfkeywords={Reactive-flow modeling, GPU, Chemical kinetics, Stiff chemistry, CUDA}}

\usepackage{booktabs}
\usepackage{multirow}
\usepackage{mathtools}
\usepackage{url, breakurl}

\def\CC{{C\nolinebreak[4]\hspace{-.05em}\raisebox{.4ex}{\tiny\bf ++}}}

\journal{Journal of Computational Physics}

\begin{document}

\begin{frontmatter}

\title{Accelerating moderately stiff chemical kinetics in reactive-flow simulations using GPUs}

\author[cwru,uconn]{Kyle~E.\ Niemeyer\corref{cor1}\fnref{padd}}
\cortext[cor1]{Corresponding author}
\ead{Kyle.Niemeyer@oregonstate.edu}
\fntext[padd]{Present address: School of Mechanical, Industrial and Manufacturing Engineering, Oregon State University, Corvallis, OR 97331}

\author[uconn]{Chih-Jen Sung}
\ead{cjsung@engr.uconn.edu}

\address[cwru]{Department of Mechanical and Aerospace Engineering \\
	Case Western Reserve University, Cleveland, OH 44106, USA}
\address[uconn]{Department of Mechanical Engineering \\
	University of Connecticut, Storrs, CT 06269, USA}

\begin{abstract}
The chemical kinetics ODEs arising from operator-split reactive-flow simulations were solved on GPUs using explicit integration algorithms. Nonstiff chemical kinetics of a hydrogen oxidation mechanism (9 species and 38 irreversible reactions) were computed using the explicit fifth-order Runge--Kutta--Cash--Karp method, and the GPU-accelerated version performed faster than single- and six-core CPU versions by factors of 126 and 25, respectively, for \num{524288} ODEs. Moderately stiff kinetics, represented with mechanisms for hydrogen\slash carbon-monoxide (13 species and 54 irreversible reactions) and methane (53 species and 634 irreversible reactions) oxidation, were computed using the stabilized explicit second-order Runge--Kutta--Chebyshev (RKC) algorithm. The GPU-based RKC implementation demonstrated an increase in performance of nearly 59 and 10 times, for problem sizes consisting of \num{262144} ODEs and larger, than the single- and six-core CPU-based RKC algorithms using the hydrogen\slash carbon-monoxide mechanism. With the methane mechanism, RKC-GPU performed more than 65 and 11 times faster, for problem sizes consisting of \num{131072} ODEs and larger, than the single- and six-core RKC-CPU versions, and up to 57 times faster than the six-core CPU-based implicit VODE algorithm on \num{65536} ODEs. In the presence of more severe stiffness, such as ethylene oxidation (111 species and \num{1566} irreversible reactions), RKC-GPU performed more than 17 times faster than RKC-CPU on six cores for \num{32768} ODEs and larger, and at best 4.5 times faster than VODE on six CPU cores for \num{65536} ODEs. With a larger time step size, RKC-GPU performed at best 2.5 times slower than six-core VODE for \num{8192} ODEs and larger. Therefore, the need for developing new strategies for integrating stiff chemistry on GPUs was discussed.
\end{abstract}


\end{frontmatter}


\section{Introduction}
\label{S:intro}

The heavy computational demands of high-fidelity computational fluid dynamics (CFD) simulations, caused by fine grid resolutions and time step sizes in addition to complex physical models, are the primary bottleneck preventing most industrial and academic researchers from performing and using such studies. Reactive-flow simulations considering detailed chemistry in particular pose prohibitive computational demands due to (1) chemical stiffness, caused by rapidly depleting species and\slash or fast reversible reactions, and (2) the large and ever-increasing size of detailed reaction mechanisms. While reaction mechanisms for fuels relevant to hypersonic engines, such as hydrogen or ethylene, may contain 10--70 species~\cite{Burke:2011fh,Qin:2000ki}, a recent surrogate mechanism for gasoline consists of about \num{1550} species and \num{6000} reactions~\cite{Mehl:2011}; a surrogate mechanism for biodiesel contains almost \num{3300} species and over \num{10000} reactions~\cite{Herbinet:2010}. Strategies for incorporating such large, realistic reaction mechanisms in reactive-flow simulations are beyond the scope of this paper; for example, Lu and Law~\cite{Lu:2009gh} recently reviewed strategies for mechanism reduction.

Even compact mechanisms pose challenges due to stiffness. In the presence of stiffness, explicit integration algorithms generally require time step sizes on the same order as the fastest chemical time scales, which can be many orders of magnitude smaller than the flow time scale~\cite{Lu:2009gh}. Due to the resulting computational inefficiency, most reactive-flow simulations rely on specialized integration algorithms such as high-order implicit solvers based on backward differentiation formulas (BDFs)~\cite{Byrne:1987wp,Hairer:2010gq}. However, these implicit solvers involve expensive linear algebra operations, so techniques for removing stiffness via reduced chemistry have also been developed~\cite{Lu:2009gh}.

Exploiting graphics processing unit (GPU) acceleration offers another avenue for enabling the use of accurate, detailed reaction mechanisms in high-fidelity reactive-flow simulations. Most reactive-flow codes rely on the operator-splitting or fractional-step method~\cite{Strang:1968wh,Knio:1999vd,Day:2000ek,Sportisse:2000gc,Oran:2001ui,Bourlioux:2003ip,Najm:2005hi,Ren:2008kd}, where the large system of governing partial differential equations (PDEs) is separated such that different physical processes are evaluated separately. For the chemistry---typically the most time-consuming portion of the simulation, accounting for 90\% or more of the total simulation time in some cases---this results in a system of independent ordinary differential equations (ODEs) for the conservation of species mass in each spatial location (i.e., at each grid point or volume).

Due to the independent nature of the integration for the systems of ODEs governing chemistry in all locations, the entire set can be integrated simultaneously. One option is to parallelize the chemistry integration on multiple central processing unit (CPU) cores or processors using the Message Passing Interface (MPI)~\cite{MPI-Forum:2009} or OpenMP~\cite{Dagum:1998hb,Chandra:2001ts,OpenMP:2008}, but the massive parallelism and increasing performance of GPUs---as well as the potential to reduce capital costs through improved energy efficiency---make them an attractive option for accelerating reactive-flow codes. General CFD applications also benefit from GPU acceleration due to the inherent data parallelism of most calculations for both finite difference and finite volume methods. Vanka et al.~\cite{Vanka:2011vc} surveyed some of the literature on using GPUs to accelerate general CFD simulations; more recently, Niemeyer and Sung~\cite{Niemeyer:2013} comprehensively reviewed advances in this area for both nonreactive and reactive flows. In the following, we will summarize important results related to GPU-based reactive-flow simulations.

The first effort in this area came from Spafford et al.~\cite{Spafford:2010ky}, who accelerated the species rate evaluations in the direct numerical simulation (DNS) code S3D~\cite{Hawkes:2005eh,Chen:2009gs} on the GPU. In their approach, the CPU handles the time integration of the chemical source terms using an explicit fourth-order Runge--Kutta method. Each integration step requires four species rate evaluations, and for each evaluation the CPU invokes the GPU to evaluate the species rates of change for all grid points simultaneously. Using an ethylene reaction mechanism with 22 species, Spafford et al.~\cite{Spafford:2010ky} achieved performance speedups of around 15$\times$ and 9$\times$ for single- and double-precision calculations, respectively.

Most recent efforts follow the spatially-independent acceleration paradigm introduced by Spafford et al.~\cite{Spafford:2010ky}, beginning with Niemeyer et al.~\cite{Niemeyer:2011uw}, who developed a GPU-based explicit integration algorithm for nonstiff chemistry. Using a compact hydrogen mechanism with 9 species and 38 irreversible reactions~\cite{Yetter:1991}, Niemeyer et al.~\cite{Niemeyer:2011uw} demonstrated a computational speedup of up to 75$\times$ compared to a single-core CPU over a wide range of independent ODE systems. Shi et al.~\cite{Shi:2012cl} presented a hybrid CPU\slash GPU chemistry integration strategy where the GPU simultaneously integrates nonstiff chemistry in grid cells using an explicit algorithm and the CPU handles spatial locations with stiff chemistry using a standard implicit integrator. This combined approach, paired with a reactive-flow code, achieved an overall performance speedup of 11--46$\times$ over the algorithms executed on a single CPU core.

Le et al.~\cite{Le:2013kt} developed the first reactive-flow solver where the GPU evaluates both the fluid transport and chemical kinetics terms. As with most other approaches, they used operator splitting to decouple and independently solve the fluid transport and chemistry terms. They handled the stiff chemical kinetics terms in parallel on the GPU using a first-order implicit method (the backward Euler method), employing a direct Gaussian elimination to solve the resulting linear system of equations. Compared against an equivalent CPU version executed on a single processor core, their combined GPU solver performed up to 40 times faster using a reaction mechanism for methane with 36 species and reversible 308 reactions, on a grid with greater than \num{e4} cells. However, the low order of the chemistry solver---first order---should be noted.

Stone et al.~\cite{Stone:2013jf} implemented two chemistry integrators on the GPU: (1) a fourth-order adaptive Runge--Kutta--Fehlberg (RKF45) method and (2) the standard fifth-order accurate implicit CVODE method. Applied to a reduced mechanism for ethylene with 19 species and 15 global reaction steps~\cite{Zambon:2007go} and compared against equivalent single-core CPU versions over a range of ODE numbers, the RKF45 and CVODE methods achieved up to 28.6$\times$ and 7.7$\times$ speedup, respectively. The GPU-based RKF45 method performed $20.2\times$ faster than the CPU-based DVODE solver operating on a single core. It should be noted that the reduced mechanism used by Stone et al.~\cite{Stone:2013jf} may not exhibit much stiffness, since it was developed by applying the quasi-steady-state approximation to certain radical species and eliminating fast elementary reactions~\cite{Zambon:2007go}.

Alternative approaches for GPU acceleration of chemical kinetics have also been presented that exploit other areas of data independence. Shi et al.~\cite{Shi:2011hv} used the GPU to (1) simultaneously calculate all the reaction rates for a single kinetic system (i.e., a single computational volume\slash grid point) and (2) accelerate the matrix inversion step of the implicit time integration using a commercial GPU linear algebra library, CULA~\cite{Humphrey:2010ga}. They found this approach beneficial for large reaction mechanisms (e.g., more than \num{1000} species), accelerating homogenous autoignition simulations up to 22$\times$, but for moderate-size mechanisms (e.g., less than 100 species) their GPU-based implementation performed slower than the original CPU version. More recently, Sankaran~\cite{Sankaran:2013he} presented a new approach for accelerating the chemistry in turbulent combustion simulations where the GPU solves the unsteady laminar flamelet equations; the controlling CPU handles the main flow solver. This method involves three levels of concurrency on the GPU: (1) the solution of species reaction rates, thermochemical properties, and molecular transport rates; (2) the solution of the discretized flamelet equations in an regular grid in the mixture fraction space; and (3) the solution of multiple flamelets.

Here, we demonstrate new strategies for accelerating chemical kinetics with moderate levels of stiffness using GPU-based explicit integration algorithms. Building upon our earlier work using the standard fourth-order Runge--Kutta algorithm~\cite{Niemeyer:2011uw}, we demonstrate the potential performance improvement using a related explicit fifth-order adaptive method for nonstiff chemical kinetics. In addition, we introduce a stabilized explicit Runge--Kutta method that can handle moderate stiffness, and show that it can be used on GPUs to achieve significant computational speedup.

The rest of the paper is structured as follows. First, we discuss some topics related to GPU computing in Section~\ref{S:gpu}. Next, in Section~\ref{S:gov-eq} we provide the governing equations for chemical kinetics in reactive-flow simulations, then in Sections~\ref{S:rkck} and \ref{S:rkc} we describe the explicit integration algorithms used in this study. In Section~\ref{S:results} we demonstrate the performance of the GPU-accelerated integration algorithms using four reaction mechanisms with increasing levels of stiffness and discuss these results. Finally, we summarize our work in Section~\ref{S:conclusions} and outline future research directions.

\section{Methodology}
\label{S:method}

\subsection{GPU computing}
\label{S:gpu}

While an in-depth discussion about GPU computing is beyond the scope of this work, we will briefly introduce important concepts. Interested readers should see the textbooks, e.g., by Kirk and Hwu~\cite{Kirk:2010we} and Sanders and Kandrot~\cite{Sanders:2010tq}. The current generation of application programming interfaces, such as CUDA~\cite{NVIDIA:2011wf} and OpenCL~\cite{Munshi:2011wk}, enables a C-like programming experience while exposing the massively parallel architecture of graphics processors. This avoids programming in the graphics pipeline directly. Our efforts are based in CUDA, a programming platform created and supported by NVIDIA, but the programming model of OpenCL, an open-source framework, is similar.

In addition, recently a new avenue for GPU parallelization has been introduced: OpenACC~\cite{OpenACC:2011vn,Reyes:2012}, which uses compiler directives (e.g., \texttt{\#pragma} statements)  placed in Fortran, C, and \CC{} codes to identify sections of code to be run in parallel on GPUs. This approach is similar to OpenMP~\cite{Dagum:1998hb,Chandra:2001ts,OpenMP:2008} for parallelizing work across multiple CPUs or CPU cores that share memory.

GPUs operate on the ``single instruction, multiple thread'' (SIMT) parallelization paradigm, similar to vector computing, where a large number of processing units independently and simultaneously execute the same instructions on different data elements. A parallel GPU function is a kernel, which---in the CUDA programming model---consists of a grid of thread blocks. Each block is made up by threads, the fundamental CUDA processing unit. Physically, GPUs consist of a number of streaming multiprocessors (e.g., 14), which can each execute 32 operations simultaneously. Thread blocks are subdivided into warps consisting of 32 threads; the streaming multiprocessors execute instructions for threads in a particular warp simultaneously. For optimal performance, all 32 threads within a warp should follow the same instruction pathway. If threads in a warp encounter different instructions (e.g., through a conditional branch), the warp diverges and significant loss in performance may result---in the worst case, by a factor of 32, if each thread follows a different instruction pathway.

\subsection{Governing equations}
\label{S:gov-eq}

Given a vector of state variables $\mathbf{\Phi} = \lbrace \phi_1, \dotsc, \phi_n \rbrace$, the governing equations for scalars in a general reactive-flow simulation are
\begin{equation}
\frac{\partial \phi_i}{\partial t} = \nabla \cdot \left( \mathbf{A}_i + \mathbf{D}_i \right) + R_i , \quad i = 1, 2, \dotsc, n,
\end{equation}
where \textbf{A} and \textbf{D} represent the advective and diffusive fluxes, respectively, and \emph{R} represents the change due to chemical reactions. Solving this stiff, coupled system for a large number of grid points\slash volumes is challenging, so many reactive-flow modeling approaches rely on operator splitting (also known as the fractional step method)~\cite{Strang:1968wh,Knio:1999vd,Day:2000ek,Sportisse:2000gc,Oran:2001ui,Bourlioux:2003ip,Najm:2005hi,Ren:2008kd}. This technique separates the integration of the stiff reaction terms from the spatially discretized transport terms, resulting in a large number of independent systems of ODEs---one for each spatial location---to solve.

When the reaction terms are separated from physical transport, the species equations are
\begin{align}
\frac{d Y_i }{dt} &= \frac{W_i \omega_i}{\rho} , \quad i = 1, 2, \dotsc, n_S , \label{E:mass} \\
\omega_i &= \sum_{j = 1}^{n_R} \left( \nu_{i j}^{\prime \prime} - \nu_{i j}^{\prime} \right) \Omega_j , \label{E:spec-rate}
\end{align}
where $Y_i$ denotes the mass fraction of the \emph{i}th chemical species, $n_S$ and $n_R$ are the numbers of species and reactions, respectively, $\rho$ is the mixture density, $W_i$ is the molecular weight of the \emph{i}th species, $\Omega_j$ is the rate of progress of reaction \emph{j}, and $\nu_{i j}^{\prime \prime}$ and $\nu_{i j}^{\prime}$ are the reverse and forward stoichiometric coefficients for the \emph{i}th species in reaction \emph{j}. The rate of progress of an irreversible reaction without pressure dependence is given by
\begin{equation}
\Omega_j = k_j \prod_{k = 1}^{n_S} C_k^{\nu_{k j}^{\prime}} , \label{E:rxn-rate}
\end{equation}
where $C_k$ is the concentration of species \emph{k}. Third-body and pressure-dependent reactions were also considered, depending on the formulation given for the particular reaction; see, for example, Law~\cite{Law:2006}, or the Chemkin manual~\cite{Kee:1996vd}, for details. The reaction rate coefficient $k_j$ follows the Arrhenius formulation
\begin{equation}
k_j = A_j T^{\beta_j} \exp \left( \frac{-E_j}{\mathcal{R} T} \right) , \label{E:rxn-const}
\end{equation}
where $\mathcal{R}$ is the universal gas constant, $A_j$ is the pre-exponential factor, $\beta_j$ is the temperature exponent, and $E_j$ is the activation energy for reaction \emph{j}. In general, reactions may be reversible; those without explicitly defined Arrhenius reverse rate parameters (i.e., \emph{A}, $\beta$, and $E$) require evaluation of the equilibrium constant to obtain their reverse rate coefficients. To avoid the conditional statements that may cause thread divergence on GPUs (as will be discussed in Section~\ref{S:results}) required by this evaluation, we converted all such reversible reactions into two irreversible reactions for each following the procedure given in \ref{A:irrev}.

In addition to the species equations, we consider a constant-pressure energy equation
\begin{equation}
\frac{dT}{dt} = -\frac{1}{\rho c_p} \sum_{i = 1}^{n_S} h_i \omega_i W_i , \label{E:energy}
\end{equation}
where $c_p$ is the mass-averaged constant-pressure specific heat and $h_i$ is the specific enthalpy of the \emph{i}th species. Together, the coupled mass and energy equations model the time-dependent behavior of an adiabatic, homogenous gas mixture in a closed system. The number of unknowns is equal to the number of species plus one (temperature), $N = n_S + 1$, and the vector of dependent variables consists of temperature and the species mass fractions, $\mathbf{y} (t) = \lbrace T, Y_1, Y_2, \dotsc, Y_{n_S} \rbrace$.

Typically, reactive-flow simulation codes use BDF-based implicit algorithms to solve Eqs.~\eqref{E:mass} and \eqref{E:energy}. While explicit algorithms tend to offer greater general efficiency and lower startup costs---important in operator-split formulations where the transport terms modify the thermochemical conditions and invalidate any saved information, such as the Jacobian matrix, between reaction integration steps---stiffness-induced instabilities force the use of extremely small time step sizes. Implicit algorithms offer greater stability and therefore allow larger time step sizes in the presence of stiffness, resulting in better performance overall. However, implicit methods involve complex control algorithms and linear algebra subroutines, with logical tests for convergence and controlling error. As such, these implicit methods may not be suitable for operating on GPUs, where the complex control flow in such operations could cause threads in a warp to diverge due to slightly different conditions. Stone et al.~\cite{Stone:2013jf} ported the implicit CVODE solver to GPU operation, and found that it performed only slightly better than a multi-core CPU version would. Explicit algorithms, on the other hand, involve simpler logical flow, and may be better-suited for GPU operation, especially with little-to-moderate stiffness in the chemical kinetics.

\subsection{Runge--Kutta--Cash--Karp method}
\label{S:rkck}

When the chemical kinetics exhibits little to no stiffness, we can solve the system of equations given by Eqs.~\eqref{E:mass} and \eqref{E:energy} using an explicit integration method such as the fifth-order Runge--Kutta method developed by Cash and Karp~\cite{Cash:1990}, namely, the RKCK method. This approach uses an embedded fourth-order method to determine the truncation error and adaptively select the step size; our methodology is taken from Press et al.~\cite{Press:1992}.

\begin{table}[tbp]
\begin{center}
\begin{tabular}{@{}c c c c c c c c c@{}}
\toprule
\emph{i} & $a_i$ & \multicolumn{5}{c}{$b_{i j}$} & $c_i$ & $c^*_i$ \\ \midrule
1 & &  & & & & & $\frac{37}{378}$ & $\frac{2825}{27648}$ \\
2 & $\frac{1}{5}$ & $\frac{1}{5}$ & & & & & 0 & 0 \\
3 & $\frac{3}{10}$ & $\frac{3}{40}$ & $\frac{9}{40}$ & & & & $\frac{250}{621}$ & $\frac{18575}{48384}$ \\
4 & $\frac{3}{5}$ & $\frac{3}{10}$ & $-\frac{9}{10}$ & $\frac{6}{5}$ & & & $\frac{125}{594}$ & $\frac{13525}{55296}$ \\
5 & 1 & $-\frac{11}{54}$ & $\frac{5}{2}$ & $-\frac{70}{27}$ & $\frac{35}{27}$ & & 0 & $\frac{277}{14336}$ \\
6 & $\frac{7}{8}$ & $\frac{1631}{55296}$ & $\frac{175}{512}$ & $\frac{575}{13824}$ & $\frac{44275}{110592}$ & $\frac{253}{4096}$ & $\frac{512}{1771}$ & $\frac{1}{4}$ \\ \midrule
\multicolumn{2}{c}{\emph{j}} & 1 & 2 & 3 & 4 & 5 & & \\
\bottomrule
\end{tabular}
\caption{Coefficients for the fifth-order Runge--Kutta--Cash--Karp method, adopted from Press et al.~\cite{Press:1992}.}
\label{T:rkck}
\end{center}
\end{table}

If $\mathbf{y}_n$ is the approximation to the exact solution $\mathbf{y}(t)$ at $t = t_n$, and $\delta t_n = t_{n+1} - t_n$ is the current step size, then the RKCK formulas, which also apply to any general fifth-order Runge--Kutta method, are
\begin{align}
\mathbf{k_1} &= \delta t \, \mathbf{f} \left( t_n, \mathbf{y}_n \right) \\
\mathbf{k}_2 &= \delta t \, \mathbf{f} \left( t_n + a_2 \, \delta t, \mathbf{y}_n + b_{2 1} \mathbf{k}_1 \right) , \\
\mathbf{k}_3 &= \delta t \, \mathbf{f} \left( t_n + a_3 \, \delta t, \mathbf{y}_n + b_{3 1} \mathbf{k}_1 + b_{3 2} \mathbf{k}_2 \right) , \\
\mathbf{k}_4 &= \delta t \, \mathbf{f} \left( t_n + a_4 \, \delta t, \mathbf{y}_n + b_{4 1} \mathbf{k}_1 + b_{4 2} \mathbf{k}_2 + b_{4 3} \mathbf{k}_3 \right) , \\
\mathbf{k}_5 &= \delta t \, \mathbf{f} \left( t_n + a_5 \, \delta t, \mathbf{y}_n + b_{5 1} \mathbf{k}_1 + b_{5 2} \mathbf{k}_2 + b_{5 3} \mathbf{k}_3 + b_{5 4} \mathbf{k}_4 \right) , \\
\mathbf{k}_6 &= \delta t \, \mathbf{f} \left( t_n + a_6 \, \delta t, \mathbf{y}_n + b_{6 1} \mathbf{k}_1 + b_{6 2} \mathbf{k}_2 + b_{6 3} \mathbf{k}_3 + b_{6 4} \mathbf{k}_4 + b_{6 5} \mathbf{k}_5 \right) , \\
\mathbf{y}_{n+1} &= \mathbf{y}_n + c_1 \mathbf{k}_1 + c_2 \mathbf{k}_2 + c_3 \mathbf{k}_3 + c_4 \mathbf{k}_4 + c_5 \mathbf{k}_5 + c_6 \mathbf{k}_6 , \\
\mathbf{y}^*_{n+1} &= \mathbf{y}_n + c^*_1 \mathbf{k}_1 + c^*_2 \mathbf{k}_2 + c^*_3 \mathbf{k}_3 + c^*_4 \mathbf{k}_4 + c^*_5 \mathbf{k}_5 + c^*_6 \mathbf{k}_6 ,
\end{align}
where $\mathbf{y}_{n+1}$ is the fifth-order solution and $\mathbf{y}^*_{n+1}$ is the solution of the embedded fourth-order method. The vector $\mathbf{f} (t, \mathbf{y} ) = d \mathbf{y} (t, \mathbf{y}) / dt $ represents the evaluation of the right-hand side of Eqs.~\eqref{E:mass} and \eqref{E:energy}. The RKCK coefficients are given in Table~\ref{T:rkck}. The fourth- and fifth-order solutions are used to estimate the error of the step $\mathbf{\Delta}_{n+1}$,
\begin{equation}
\mathbf{\Delta}_{n+1} = \mathbf{y}_{n+1} - \mathbf{y}^*_{n+1} = \sum_{i = 1}^6 \left( c_i - c^*_i \right) \mathbf{k}_i .
\end{equation}
This error is then compared against a desired accuracy, $\mathbf{\Delta_0}$, defined by
\begin{equation}
\mathbf{\Delta}_0 = \epsilon \left( | \mathbf{y}_n | + \left| \delta t \, \mathbf{f} \left(t_n, \mathbf{y}_n \right) \right| + \delta \right) ,
\end{equation}
where $\epsilon$ is a tolerance level and $\delta$ represents a small value (e.g., \num{e-30}). If the estimated error of the current step is larger than the desired accuracy ($ \mathbf{\Delta}_{n+1} > \mathbf{\Delta}_0 $), the step is rejected and a smaller step size is calculated; if the error is smaller than the desired accuracy ($ \mathbf{\Delta}_{n+1} \leq \mathbf{\Delta}_0 $), the step is accepted and the step size for the next step is calculated. The following is used to calculate a new step size based on error and the current step size:
\begin{equation}
\delta t_{\text{new}} =
\begin{dcases}
S \, \delta t_n \, \max_i \left( \left| \frac{\Delta_{0,i}}{\Delta_{n+1, i}} \right| \right)^{1/5} \quad \text{if } \mathbf{\Delta_{n+1}} \leq \mathbf{\Delta}_0 , \\
S \, \delta t_n \, \max_i \left( \left| \frac{\Delta_{0,i}}{\Delta_{n+1, i}} \right| \right)^{1/4} \quad \text{if } \mathbf{\Delta_{n+1}} > \mathbf{\Delta}_0 .
\end{dcases}
\label{E:hnew}
\end{equation}
Here, \emph{i} represents the \emph{i}th element of the related vector and \emph{S} denotes a safety factor slightly smaller than unity (e.g., 0.9). Eq.~\eqref{E:hnew} is used to calculate the next time step size for an accepted step and a new, smaller step size when the error is too large (and therefore the step is rejected). In practice, step size decreases and increases are limited to factors of ten and five, respectively.

\subsection{Runge--Kutta--Chebyshev method}
\label{S:rkc}

For stiff problems, standard explicit integration methods become unsuitable due to stability issues, requiring unreasonably small time step sizes~\cite{Hairer:2010gq}. Traditionally, implicit integration algorithms such as those based on BDFs have been used to handle stiff problems, but these require expensive linear algebra operations on the Jacobian matrix. In addition, the complex logical flow would result in highly divergent instructions across different initial conditions, making implicit algorithms unsuitable for operation on GPUs. One alternative to implicit algorithms for problems with moderate levels of stiffness is a stabilized explicit scheme such as the Runge--Kutta--Chebyshev (RKC) method~\cite{Houwen:1980,Verwer:1990tg,vanderHouwen:1996ti,Verwer:1996vo,Sommeijer:1997uv,Verwer:2004gf}. While the RKC method is explicit, it is capable of handling stiffness through additional stages---past the first two required for second-order accuracy---that extend its stability domain along the negative real axis.

Our RKC implementation is taken from Sommeijer et al.~\cite{Sommeijer:1997uv} and Verwer et al.~\cite{Verwer:2004gf}. Following the same terminology as in the description of the RKCK method in Section~\ref{S:rkck}, where $\mathbf{y}_n$ is the approximation to the exact solution $\mathbf{y}(t)$ at $t = t_n$ and $\delta t_n = t_{n+1} - t_n$ is the current step size, the formulas for the second-order RKC are
{\allowdisplaybreaks \begin{IEEEeqnarray}{rCl}
\mathbf{w}_0 & = & \mathbf{y}_n , \label{E:rkc0} \\
\mathbf{w}_1 & = & \mathbf{w}_0 + \tilde{\mu}_1 \, \delta t \, \mathbf{f}_0 , \label{E:rkc1} \\
\mathbf{w}_j & = & (1 - \mu_j - \nu_j ) \mathbf{w}_0 + \mu_j \mathbf{w}_{j - 1} \nonumber \\
& & +\: \nu_j \mathbf{w}_{j - 2} + \tilde{\mu}_j \, \delta t \, \mathbf{f}_{j - 1} + \tilde{\gamma}_j \, \delta t \, \mathbf{f}_0, \quad j = 2, \dotsc, s ,  \label{E:rkcj} \\
\mathbf{y}_{n + 1} & = & \mathbf{w}_s , \label{E:rkcs}
\end{IEEEeqnarray}}%
where \emph{s} is the total number of stages. The $\mathbf{w}_j$ are internal vectors for the stages, and $\mathbf{f}_j$ are evaluations of the right-hand-side function of the governing equations at each stage, where $\mathbf{f}_j = \mathbf{f} (t_n + c_j \, \delta t , \mathbf{w}_j )$. Note the recursive nature of $\mathbf{w}_j$, which requires only five arrays for storage. The coefficients used in Eqs.~\eqref{E:rkc1} and \eqref{E:rkcj} are available analytically for any $s \geq 2$:
\begin{align}
\tilde{\mu}_1 &= b_1 \omega_1 , \\
\mu_j = \frac{2 b_j \omega_0}{b_{j - 1}} , \quad \nu_j &= \frac{-b_j}{b_{j-2}}, \quad \tilde{\mu}_j = \frac{2 b_j \omega_1}{b_{j-1}}, \quad \tilde{\gamma}_j = -a_{j-1} \tilde{\mu_j} \\
b_0 = b_2, \quad b_1 &= \frac{1}{\omega_0}, \quad b_j = \frac{T_j^{\prime \prime} (\omega_0) }{ \left( T_j^{\prime} (\omega_0) \right)^2 }, \\
w_0 = 1 + \frac{\kappa}{s^2} , \quad \omega_1 &= \frac{ T_s^{\prime} (\omega_0) }{ T_s^{\prime \prime} (\omega_0) } ,
\end{align}
where $\kappa \geq 0$ is the damping parameter (we used $\kappa = 2 / 13$~\cite{Sommeijer:1997uv,Verwer:2004gf}). $T_j(x)$ are the Chebyshev polynomials of the first kind, defined recursively as
\begin{equation}
T_j (x) = 2 x T_{j - 1} (x) - T_{j - 2} (x), \quad j = 2, \dotsc, s,
\end{equation}
where $T_0 (x) = 1$, $T_1 (x) = x$, and $T_j^{\prime}(x)$ and $T_j^{\prime \prime}(x)$ are the first and second derivatives of $T_j (x)$, respectively. The $c_j$ used in the function evaluations are
\begin{align}
c_1 &= \frac{c_2}{T_2^{\prime}(\omega_0)} \approx \frac{c_2}{4}, \\
c_j &= \frac{T_s^{\prime} (\omega_0)}{T_s^{\prime \prime} (\omega_0)} \frac{T_j^{\prime \prime} (\omega_0)}{T_j^{\prime} (\omega_0)} \approx \frac{j^2 - 1}{s^2 - 1}, \quad 2 \leq j \leq s - 1, \\
c_s &= 1.
\end{align}

The RKC method can also be used with an adaptive time stepping method for error control, as given by Sommeijer et al.~\cite{Sommeijer:1997uv}. After taking the step $t_{n+1} = t_n + \delta t_n$ and calculating $\mathbf{y}_{n+1}$, the error in the calculation at the current step is estimated using
\begin{equation}
\mathbf{\Delta}_{n+1} = \frac{4}{5} (\mathbf{y}_n - \mathbf{y}_{n+1}) + \frac{2}{5} \delta t_n (\mathbf{f}_n + \mathbf{f}_{n+1}) .
\end{equation}
These error estimates are used with absolute and relative tolerances to define the weighted RMS norm of error:
\begin{align}
\| \mathbf{\Delta}_{n+1} \|_{\text{rms}} &= \left \| \frac{\mathbf{\Delta}_{n+1}}{\mathbf{T} \sqrt{N}} \right \|_2 , \label{E:rmsE} \\
\mathbf{T} &= \mathbf{A} + R \cdot \max \left( | \mathbf{y}_n |, | \mathbf{y}_{n+1} | \right) ,
\end{align}
where \emph{N} represents the number of unknown variables (here, $N = n_S + 1$ as defined previously), $\mathbf{A}$ is the vector of absolute tolerances, and \emph{R} is the relative tolerance. The norm $\| \cdot \|_2$ indicates the Euclidean or $L_2$ norm. The step is accepted if $ \| \mathbf{\Delta}_{n+1} \|_{\text{rms}} \leq 1 $; otherwise, it is rejected and redone using a smaller step size. The weighted RMS norm of error for the current and prior steps, and the associated step sizes, are then used to predict the new step size, using
\begin{align}
\delta t_{n+1} &= \min \left( 10, \max( 0.1, f ) \right) \delta t_n , \\
f &= 0.8 \left( \frac{ \| \mathbf{\Delta}_n \|_{\text{rms}}^{1 / (p + 1)} }{ \| \mathbf{\Delta}_{n+1} \|_{\text{rms}}^{1 / (p + 1)} } \frac{\delta t_n}{\delta t_{n - 1}} \right) \frac{1}{ \| \mathbf{\Delta}_n \|_{\text{rms}}^{1 / (p + 1)} } ,
\end{align}
where \emph{p} is the order of the algorithm---two, in this case. When a step is rejected, we use a similar equation to calculate a new step size:
\begin{equation}
f = \frac{0.8}{ \| \mathbf{\Delta}_n \|_{\text{rms}}^{1 / (p + 1)} } .
\end{equation}

In order to determine the initial time step size, we first use a tentative step size calculated as the inverse of the spectral radius $\sigma$---the magnitude of the largest eigenvalue---of the Jacobian. After predicting the error associated with this tentative step, we then set the initial step size as one-tenth of the step size that would satisfy error control based on the tentative step:
\begin{align}
\delta t_0 &= \frac{1}{ \sigma } , \\
\mathbf{\Delta}_0 &= \delta t_0 \left( \mathbf{f}(t_0 + \delta t_0, \mathbf{y}_0 + \delta t_0 \, \mathbf{f}(t_0, \mathbf{y}_0)) - \mathbf{f} (t_0, \mathbf{y}_0) \right), \\
\delta t_1 &= 0.1 \frac{\delta t_0}{ \| \mathbf{\Delta_0} \|_{\text{rms}}^{1/2} } ,
\end{align}
where $\|\mathbf{\Delta_0}\|_{\text{rms}}$ is evaluated in the same manner as $\| \mathbf{\Delta}_{n+1} \|_{\text{rms}}$ using Eq.~\eqref{E:rmsE}.

After selecting the optimal time step size to control local error, the algorithm then determines the optimal number of RKC stages in order to remain stable. Due to stiffness, too few stages would lead to instability. The local stiffness is determined using the spectral radius and time step size. The number of stages are determined by
\begin{equation}
s = 1 + \sqrt{1 + 1.54 \, \delta t_n \, \sigma} ,
\end{equation}
as suggested by Sommeijer et al.~\cite{Sommeijer:1997uv}, where the value 1.54 is related to the stability boundary of the algorithm. Note that \emph{s} may vary between time steps due to a changing spectral radius and time step size. In our RKC implementation, we used a nonlinear power method~\cite{Sommeijer:1997uv} to calculate the spectral radius; this choice costs an additional vector to store the computed eigenvector, but avoids storing or calculating the Jacobian matrix. Depending on the problem type, alternative methods such as the Gershgorin circle theorem~\cite{Gersgorin:1931,Horn:1990} could be used to obtain an upper-bound estimate for the spectral radius. In our experience, however, the circle theorem tended to overestimate the spectral radius, resulting in unnecessarily large numbers of stages---this induced greater computational expense compared to using the power method. Following Sommeijer et al.~\cite{Sommeijer:1997uv}, in our RKC implementation the spectral radius is estimated every 25 (internal) steps or after a step rejection. In addition, the computed eigenvector is saved to be used as the initial guess in the next evaluation.

\section{Results and discussion}
\label{S:results}

In order to study the performance of the GPU-based RKCK and RKC solvers (termed RKCK-GPU and RKC-GPU, respectively), we tested their performance with four reaction mechanisms, representing different levels of stiffness. We varied the problem size, meaning number of chemical kinetics ODEs, over a wide range from \num{e2} to \num{e6}, representing a wide range of grid resolutions in an operator-split reactive-flow code.

First, we studied the performance of RKCK-GPU using a nonstiff hydrogen mechanism. Next, we considered (separately) mechanisms for hydrogen\slash carbon monoxide and methane with moderate levels of stiffness and use these to study the performance of RKC-GPU. Finally, we examined the performance of RKC-GPU in a case where stiffness is more severe, using an ethylene mechanism. In all four cases, we compared the performance of the GPU algorithm against an equivalent CPU version. In the presence of stiffness, we also compared the performance of RKC-GPU against an implicit CPU-based code, VODE\_F90~\cite{Byrne:2006}, a Fortran 90 version of the well-known VODE solver.

In both the CPU and GPU algorithms used here, we generated the subroutines needed for chemical kinetics source terms (e.g., species rates, reaction rates, thermodynamic properties) using an open-source Python tool that we created~\cite{Niemeyer:2013cs}, which takes Chemkin-format reaction mechanisms as input. Further, we converted all reversible reactions in the reaction mechanisms used here into two irreversible reactions for each in order to avoid the computation of equilibrium constants, as described in \ref{A:irrev}. We developed an additional Python tool implementing this procedure that is also available online~\cite{Niemeyer:2013im}. We paired VODE with CHEMKIN-III~\cite{Kee:1996vd} to evaluate the chemical kinetics and species thermodynamic properties. All calculations were performed in double precision and at constant pressure---although the generated subroutines are also capable of constant volume conditions.

All calculations reported here were performed using a single GPU and single CPU; we measured the serial CPU performance using a single core as well as parallelized CPU performance---via OpenMP~\cite{OpenMP:2008}---on six cores. The GPU calculations were performed using an NVIDIA Tesla c2075 GPU with \SI{6}{\giga\byte} of global memory. An Intel Xeon X5650 CPU, running at \SI{2.67}{\giga\hertz} with \SI{256}{\kilo\byte} of L2 cache memory per core and \SI{12}{\mega\byte} of L3 cache memory, served as the host processor for the GPU calculations and ran the CPU single- and six-core OpenMP calculations. We used the GNU Compiler Collection (gcc) version 4.6.2 (with the compiler options ``\texttt{-O3 -ffast-math -std=c99 -m64}'') to compile the CPU programs and the CUDA 5.0 compiler nvcc version 0.2.1221 (``\texttt{-O3 -arch=sm\_20 -m64}'') to compile the GPU versions. The function \texttt{cudaSetDevice()} was used to hide any device initialization delay in the CUDA implementations prior to the timing.

Imposing identical initial conditions for all ODEs would not represent the situation in a reactive-flow simulation where conditions vary across space, so we generated initial conditions for the ODEs by sampling the solutions obtained from constant pressure homogeneous ignition simulations. For all four fuels studied, we used starting conditions of \SI{1600}{\kelvin}, \SI{1}{\atm}, and an equivalence ratio of one. This resulted in a set of initial conditions covering a wide range of temperatures and species mass fractions. For example, some data points came from the pre-ignition induction period, some from the transient regime when temperature increases rapidly, and some from the post-ignition stage where conditions approach equilibrium. We distributed the resulting initial conditions in two ways. First, we assigned initial conditions sequentially to ODEs, where consecutive data points---taken from consecutive time steps---contain similar conditions. This emulated adjacent spatial locations with similar but not identical conditions. Further, this procedure represents a more realistic performance measure compared to the previous work of Niemeyer et al.~\cite{Niemeyer:2011uw}, where identical initial conditions and a constant time step size were used. For the GPU-based algorithms, similar---but not identical---initial conditions will result in threads within warps that may follow divergent pathways, due to varying time step sizes, for example. In order to further explore the impact of divergence on performance, we also assigned initial conditions to threads in a second manner: randomly shuffling the order. Compared to using similar conditions, randomly selected initial conditions represent a worst-case potential for divergence.

Other potential sources of thread divergence could be conditional statements in the source terms, because, e.g., thermodynamic properties are typically fitted as polynomials across different temperature ranges, certain reaction pressure-dependence formulations are described in different pressure ranges. We attempted to minimize the occurrence of such conditional statements by converting each reversible reaction in the reaction mechanisms into a pair of irreversible reactions (as described above). This avoided the temperature conditional statement required for evaluating the Gibbs function polynomial, in turn needed for the equilibrium constants. Regarding the conditional statements required to evaluate the species thermodynamic properties for the energy equation or reaction rates for particular pressure-dependence formulations, in the current work, neither of these contributed to thread divergence because (1) all temperatures experienced by threads fell within the same polynomial fitting range and (2) none of the pressure-dependent reactions considered in the reaction mechanisms were formulated using multiple pressure ranges. However, in general cases, conditional statements on temperature or pressure could cause additional thread divergence.

The integration algorithms take as input initial conditions and a global time step, performing internal sub-stepping as necessary. The computational times, or wall-clock times, reported represent the average over ten global time steps. For the GPU implementations, the reported computational time per global time step included the overhead required for transmitting data between the CPU and GPU before and after each integration step. The integrator restarts at each global time step,  not storing any data from the previous step---although any sub-stepping performed by the algorithm within these larger steps does benefit from retained information from prior sub-steps. This is done to emulate a true operator-split code, where the transport integration step would update the thermochemical conditions independently from the chemistry and therefore invalidate any retained information between global steps. This reduces the efficiency somewhat, by forcing the integrator to take initially large test steps, but the startup costs of the explicit integration algorithms considered here pale in comparison to those of implicit integrators such as VODE, where the Jacobian matrix must be re-evaluated.

In the GPU-based algorithms, threads independently integrated each chemical kinetics ODE. The total number of threads then equaled the number of ODEs; blocks consisted of 64 threads each. For problem sizes of \num{4194304} or larger, where a block size of 64 threads would exceed the maximum limit on number of blocks per grid (\num{65535}) in one dimension, we used a block size of $N_t / \num{32768}$, where $N_t$ is the total number of threads. We kept the block size as a multiple of 32 to ensure blocks contained whole thread warps.

\subsection{Hydrogen kinetics}
\label{S:h2}

First, we considered a case where stiffness in the chemical kinetics does not pose a challenge, using the hydrogen oxidation mechanism of Yetter et al.~\cite{Yetter:1991} with 9 species and 38 irreversible reactions. We employed the explicit RKCK method, with a tolerance level $\epsilon$ of \num{1e-10}, and performed 10 global integration steps of \SI{1e-8}{\second} (or \SI{10}{\nano\second}) each. The average time needed per step is reported. This application is relevant particularly for DNS and studies of high-speed flows, which use extremely short time step sizes in order to resolve the Kolmogorov scales and capture the short time scales due to high flow velocity, respectively. Adjacent ODEs used similar initial conditions as described in the previous section.

The lack of stiffness in this case was due to both the particular chemistry considered and the short global time step sizes used. Quantifying stiffness is somewhat difficult~\cite{Hairer:2010gq}, but in general explicit methods are more efficient for nonstiff problems than implicit or other stiff integrators (e.g., stabilized explicit methods like RKC). In terms of computational time, RKCK-GPU performed nearly 2.3$\times$ and 2.6$\times$ faster than RKC-GPU for problem sizes of \num{65536} and \num{262144} independent ODEs, respectively, so we consider this case nonstiff.

\begin{figure}[tbp] \begin{center}
\includegraphics[width=0.8\linewidth]{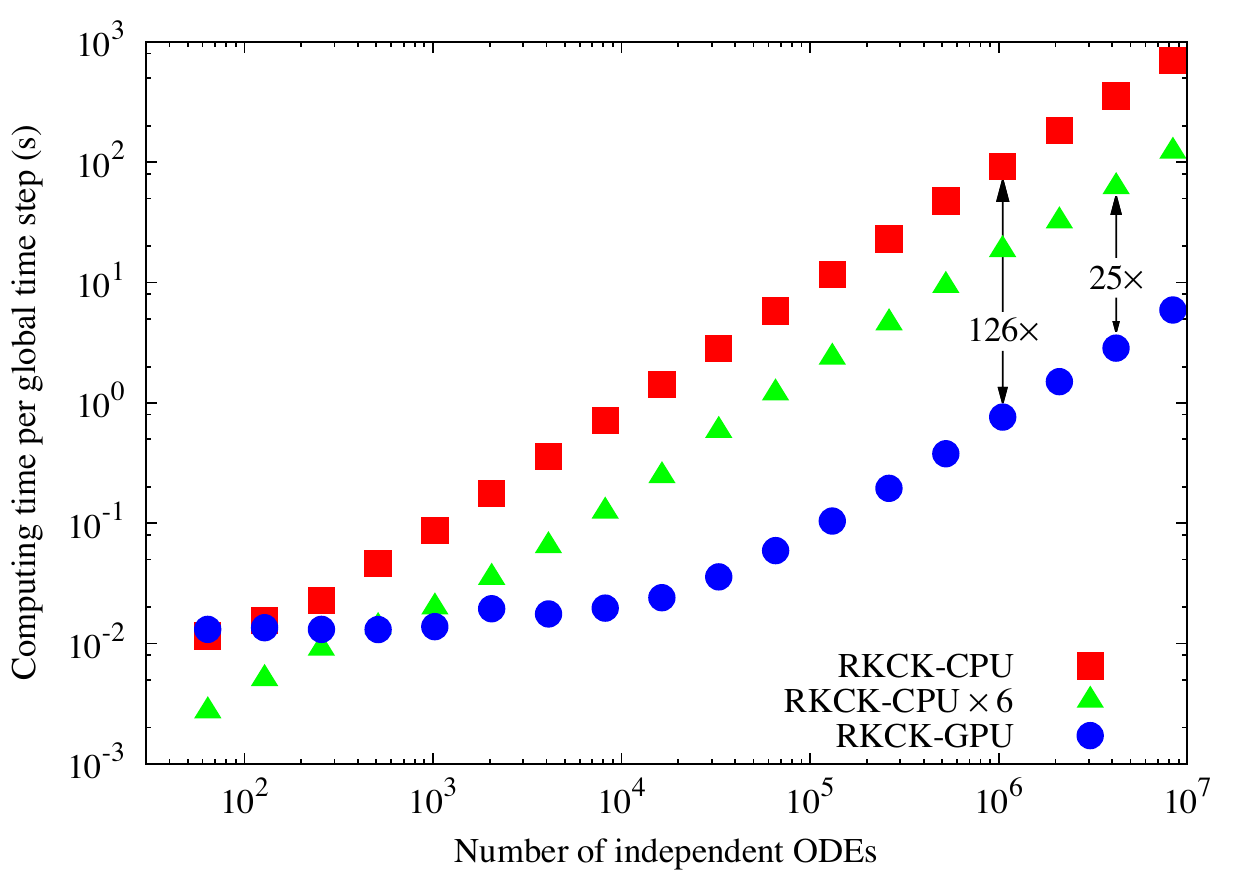}
\caption{Performance comparison of (single- and six-core) CPU and GPU integration of the nonstiff hydrogen mechanism using the explicit RKCK method. Note that both axes are displayed in logarithmic scale.}
\label{F:h2-rkck}
\end{center} \end{figure}

Figure~\ref{F:h2-rkck} shows the performance results of the CPU- and GPU-based RKCK algorithms for problem sizes ranging from 64 to \num{8388608}. RKCK-GPU performed faster than the single-core RKCK-CPU for problem sizes of 128 ODEs and larger, and faster than the six-core CPU version when the number of ODEs is 512 or larger. Note that the speedup of the GPU implementation increased with growing problem size. For the largest problem sizes, RKCK-GPU ran up to 126$\times$ and 25$\times$ faster than the single- and six-core RKCK-CPU versions. On six cores RKCK-CPU ran between five and six times faster than on a single core, due to the data independent nature of the problem.

\begin{figure}[tbp]
\begin{center}
\includegraphics[width=0.8\linewidth]{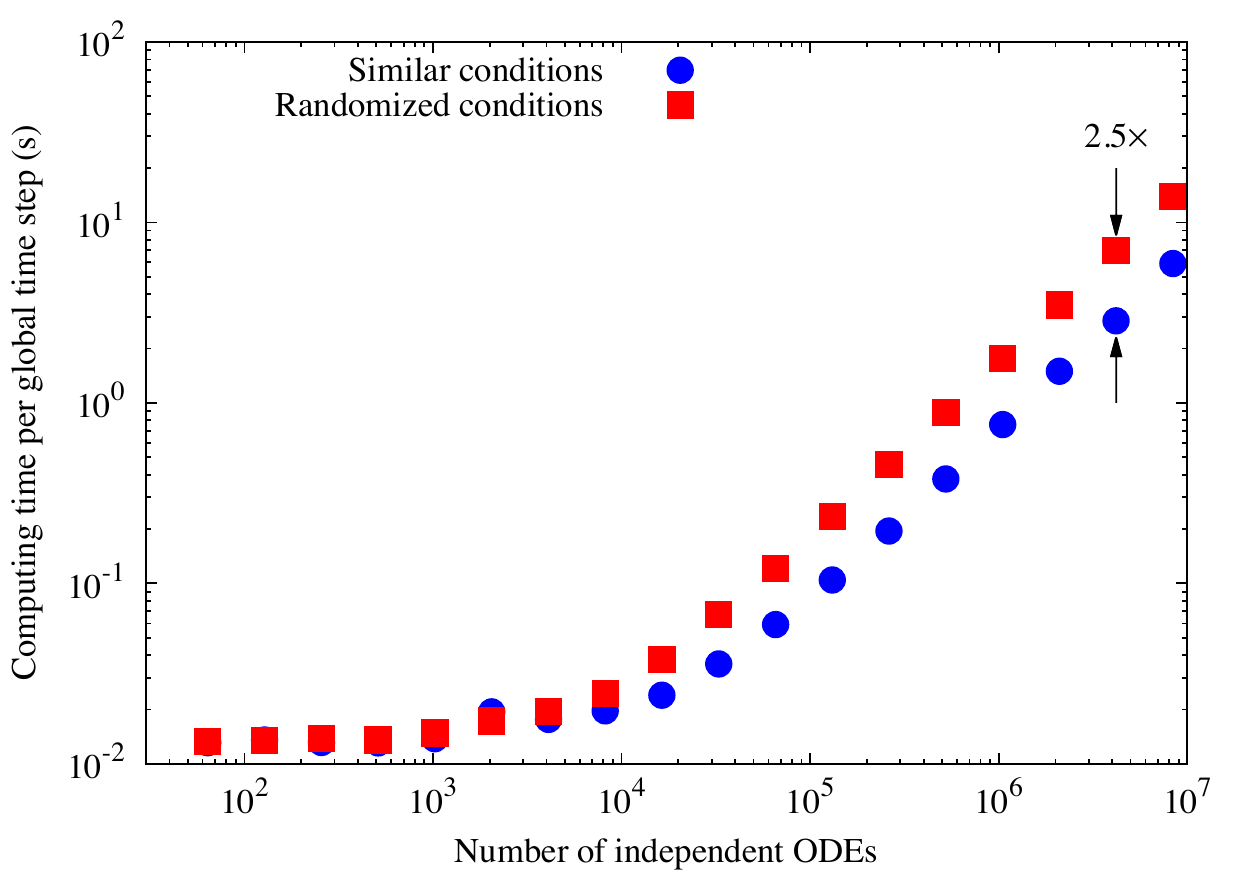}
\caption{Performance comparison of RKCK-GPU integration of the nonstiff hydrogen mechanism where neighboring threads have similar and randomized initial conditions.}
\label{F:h2-rkck-random}
\end{center}
\end{figure}

We also studied the effect of different initial conditions on the performance of RKCK-GPU, by randomly shuffling the data points used for this purpose such that neighboring threads no longer contained similar data. This resulted in thread divergence, since different threads in each warp will require different inner time step sizes---therefore some threads will require a greater number of steps, while others will finish sooner. Figure~\ref{F:h2-rkck-random} shows the comparison of performance for RKCK-GPU between threads with similar initial conditions and threads where initial conditions were randomly selected (and are therefore different). The divergence caused by randomized initial conditions reduced performance by up to a factor of 2.3, with a greater reduction at larger problem sizes. We note that some divergence was also present for threads with similar---but not identical---initial conditions.

\begin{figure}[tbp]
\begin{center}
\includegraphics[width=.8\linewidth]{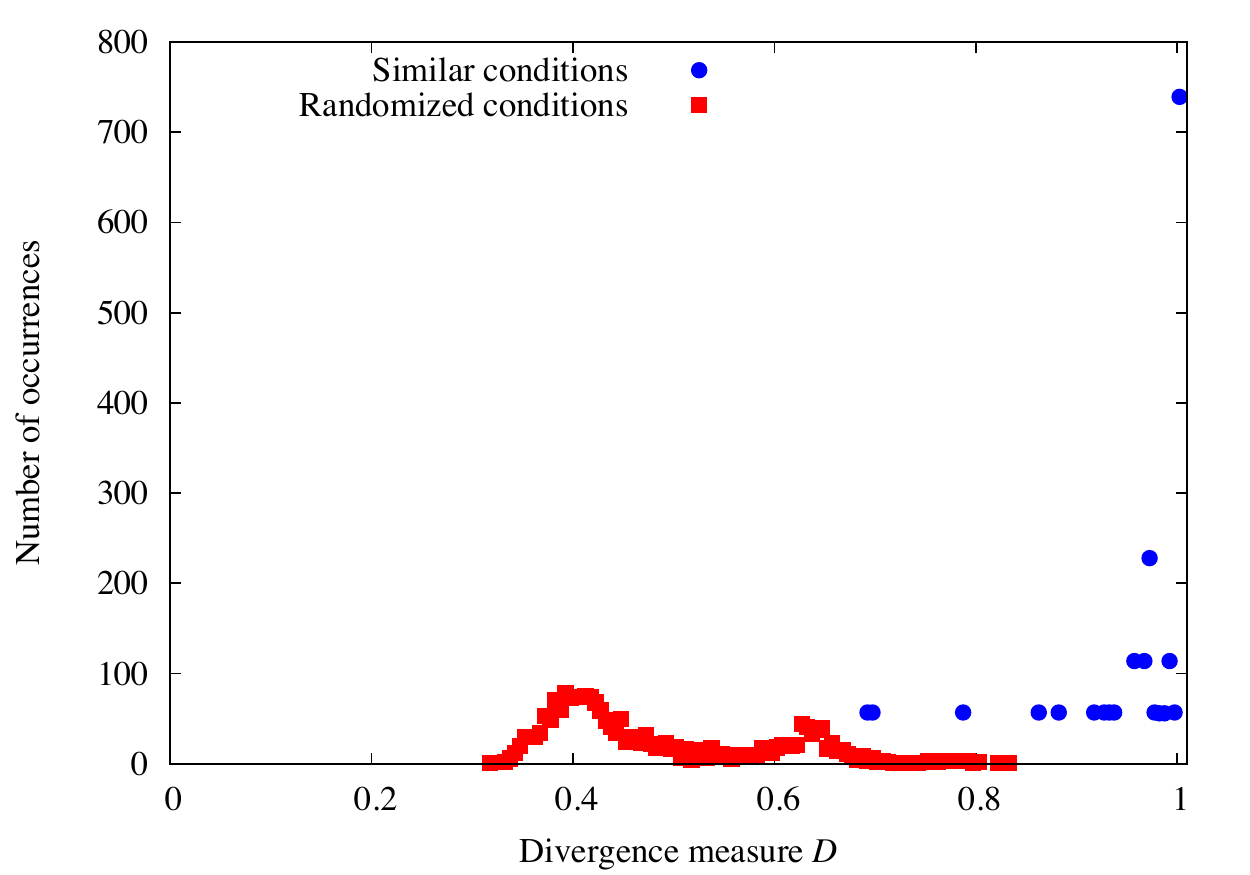}
\caption{Warp thread divergence comparison of RKCK-GPU for nonstiff hydrogen kinetics for similar and random initial conditions, where the number of occurrences of the divergence measure \emph{D} is plotted for \num{2048} thread warps.}
\label{F:h2-rkck-diverge}
\end{center}
\end{figure}

In order to quantify this divergence, we introduce a measure for the divergence in a thread warp, \emph{D}, proposed by Stone~\cite{Stone:2013} and Sankaran~\cite{Sankaran:2013}, defined by
\begin{equation}
D = \frac{ \sum_{i = 1}^{32} d_i }{32 \, \max_{i} d_i } ,
\end{equation}
where $d_i$ denotes the number of right-hand function (i.e., derivative) evaluations in the \emph{i}th thread over a certain number of global time steps. We used this to represent the cost of integration per global step for each thread within a warp. Values of \emph{D} approaching one represent a warp with completely converged threads, while values approaching zero represent a situation where a small number of threads perform significantly more work than other threads. However, it should be noted that \emph{D} is not a perfect measure of divergence in general applications, where threads may follow different instructions but perform similar amounts of work. Figure~\ref{F:h2-rkck-diverge} shows the distribution of \emph{D} for \num{65536} ODEs, corresponding to \num{2048} thread warps, where the sum of the derivative evaluations over ten global time steps was used to evaluate \emph{D}. For similar initial conditions, the divergence remained low as measured by \emph{D}, while with randomized initial conditions the divergence was greater, with \emph{D} ranging between 0.3--0.8 and showing peaks around 0.4 and 0.65. This divergence likely caused the reduced performance of RKCK-GPU with randomly chosen initial conditions compared to similar initial conditions.

\subsection{Hydrogen\slash carbon monoxide kinetics}
\label{S:h2-co}

Next, we studied a kinetic system with moderate stiffness, using the hydrogen\slash carbon monoxide reaction mechanism of Burke et al.~\cite{Burke:2011fh}, which consists of 13 species and 27 reversible (converted to 54 irreversible) reactions. Here, we chose a global time step size of \SI{1e-6}{\second} and reported the average computational time for ten steps. This value represents step sizes used in large-eddy simulations of reactive flows~\cite{Wang:2011kq,Bulat:2013ds}. We consider this problem to be ``moderately'' stiff because RKC-GPU performed more than three times faster than RKCK-GPU. For RKC, we used a relative tolerance of \num{1e-6} and an absolute tolerance of \num{1e-10}. Adjacent ODEs used similar initial conditions.

\begin{figure}[tbp]
\begin{center}
\includegraphics[width=0.8\linewidth]{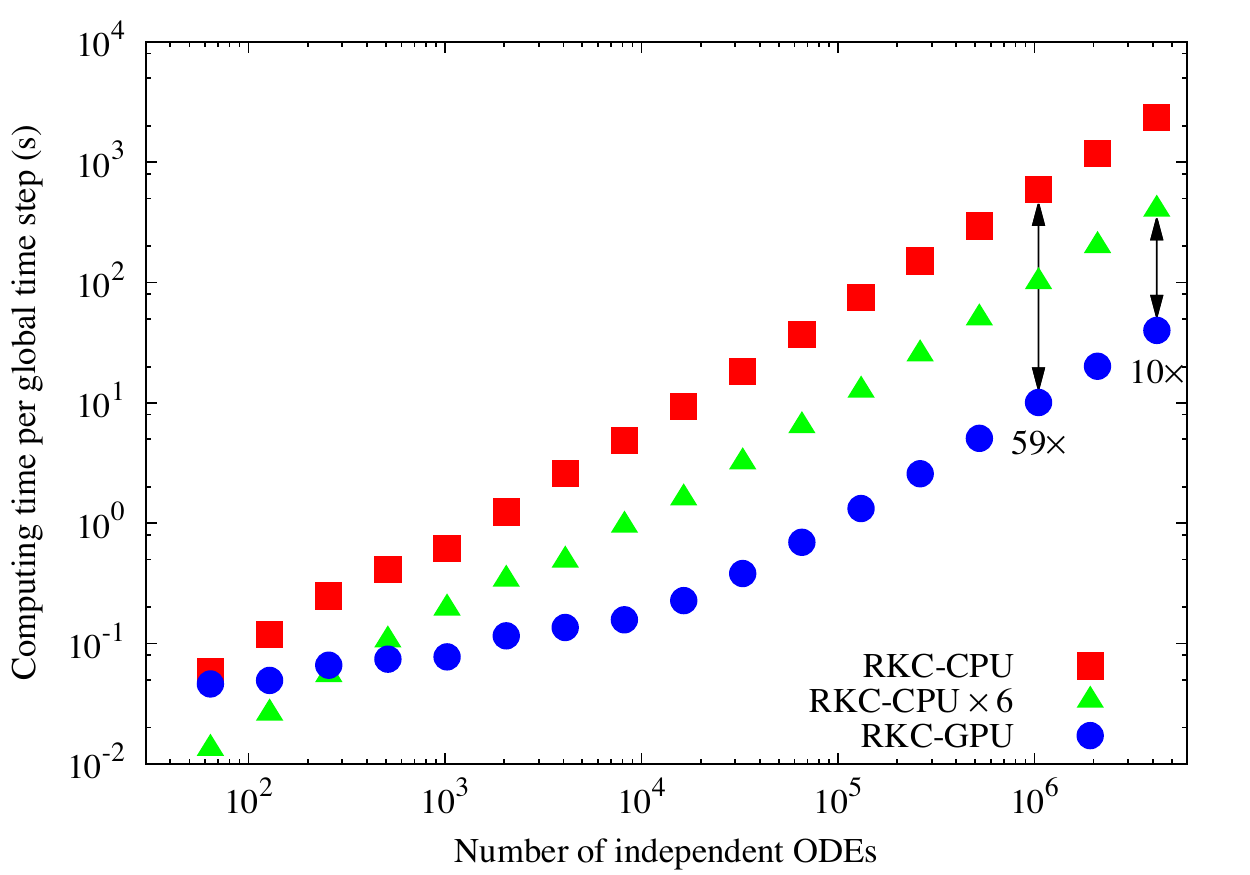}
\caption{Performance comparison of (single- and six-core) CPU and GPU integration of the moderately stiff hydrogen\slash carbon monoxide mechanism using the stabilized explicit RKC method.}
\label{F:h2-co-rkc}
\end{center}
\end{figure}

Figure~\ref{F:h2-co-rkc} shows the performance comparison between the single- and six-core RKC-CPU and RKC-GPU for problem sizes ranging from 64 to \num{4194304}. As with the RKCK algorithm, at smaller problem sizes RKC-GPU compared less favorably against RKC-CPU, but the speedup increased with increasing problem size. RKC-GPU outperformed RKC-CPU on a single CPU core for the entire range of ODE numbers considered here, while it performed faster than the six-core version for problem sizes of 512 ODEs and larger. While the exact speedup varied, for ODE numbers of \num{262144} and higher RKC-GPU demonstrated performance speedups of 59$\times$ and 10$\times$ compared to RKC-CPU on one and six CPU cores, respectively.

\begin{figure}[tbp] \begin{center}
\includegraphics[width=0.8\linewidth]{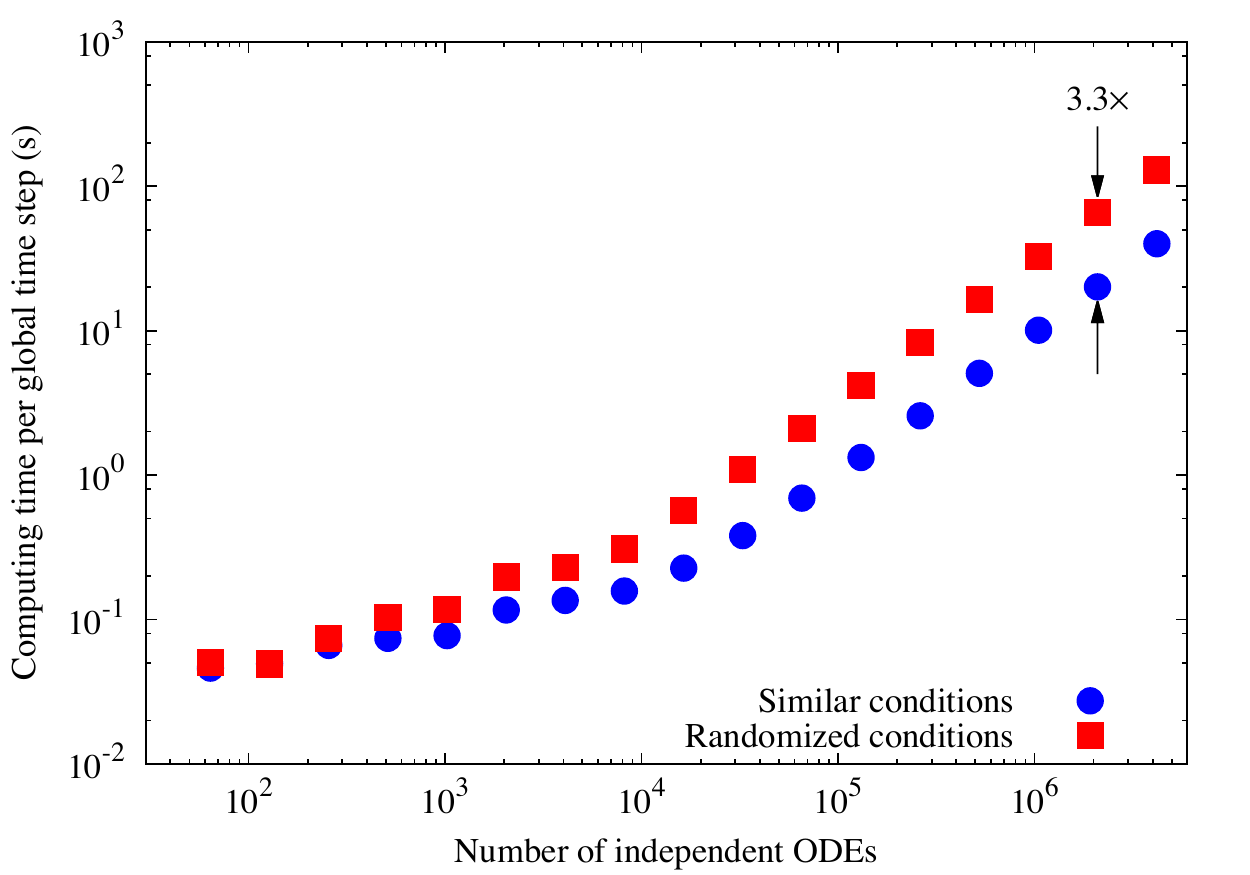}
\caption{Performance comparison of RKC-GPU integration of the hydrogen\slash carbon monoxide mechanism where neighboring threads have similar and randomized initial conditions.}
\label{F:h2-rkc-random}
\end{center} \end{figure}

Similar to our analysis of divergence for RKCK-GPU, we also studied the effect of randomized initial conditions on the performance of RKC-GPU. In this case, there are now three potential sources of thread divergence: (1) varying numbers of iterations for the nonlinear power method used to estimate the spectral radius, (2) varying numbers of stages due to different spectral radii, and (3) varying numbers of steps due to different time step sizes. Figure~\ref{F:h2-rkc-random} shows the performance comparison for RKC-GPU between threads with similar and randomized initial conditions. Thread divergence caused by random initial conditions reduced the performance of RKC-GPU by up to a factor of 3.3. As expected, RKC-GPU exhibited a greater performance loss than RKCK-GPU, where the major source of thread divergence was varying numbers of time steps.

\begin{figure}[tbp]
\begin{center}
\includegraphics[width=.8\linewidth]{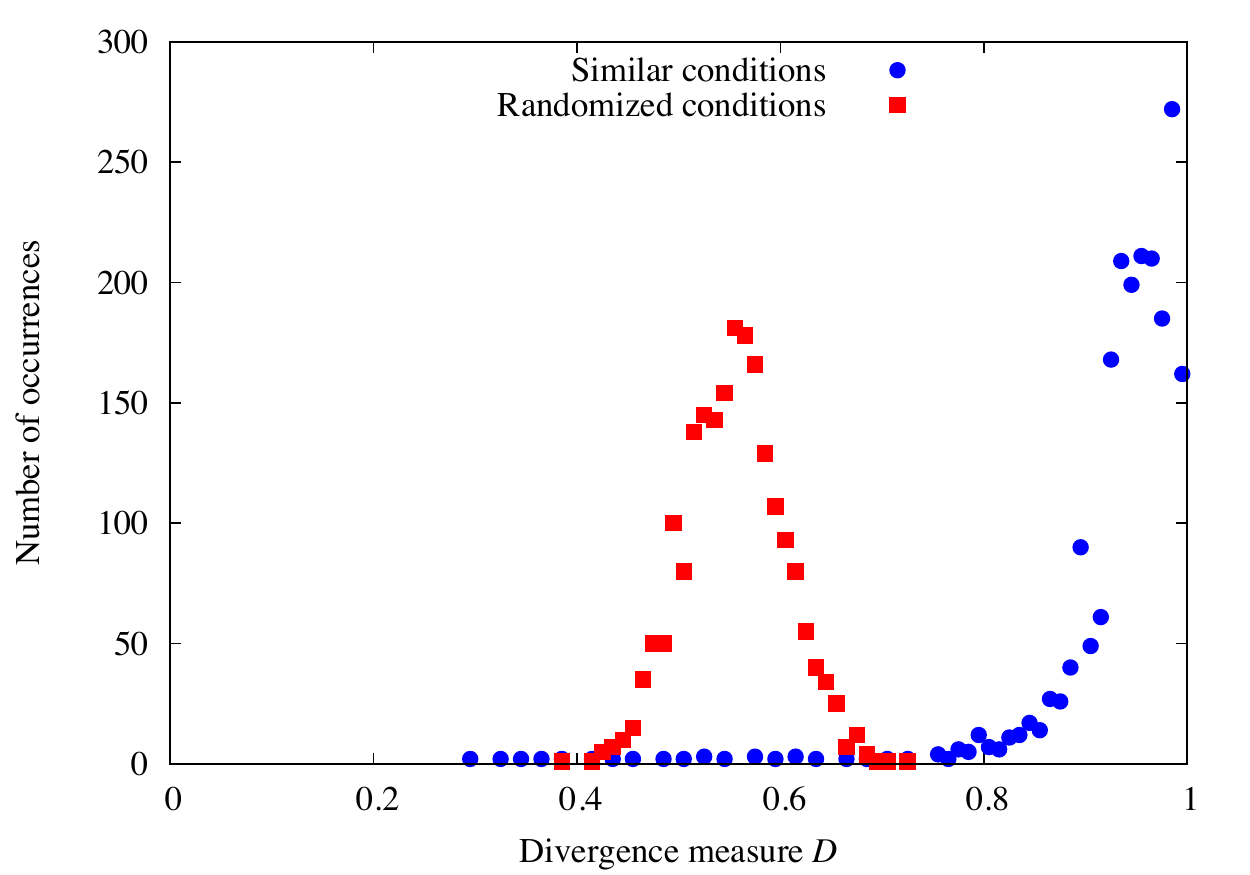}
\caption{Warp thread divergence comparison of RKC-GPU for hydrogen\slash carbon monoxide kinetics for similar and random initial conditions, where the number of occurrences of the divergence measure \emph{D} is plotted for \num{2048} thread warps.}
\label{F:h2-rkc-diverge}
\end{center}
\end{figure}

The greater divergence of RKC-GPU is also demonstrated in Fig.~\ref{F:h2-rkc-diverge}, where the number of occurrences of \emph{D} is counted for \num{65536} ODEs (\num{2048} warps). In this case, even similar initial conditions caused some divergence. This was likely the reason for the reduced performance speedup of RKC-GPU compared to that of RKCK-GPU relative to their respective CPU versions. With randomly distributed initial conditions, \emph{D} is distributed normally around $\sim$0.55. Compared to the distribution of \emph{D} for RKCK, RKC shows a higher incidence of low values, likely the cause behind the greater reduction in performance for the randomized initial condition case with RKC.

\subsection{Methane kinetics}
\label{S:methane}

Next, we analyzed the performance of the CPU and GPU versions of RKC in another case with moderate stiffness, using the GRI-Mech 3.0~\cite{Smith:2010} mechanism for methane oxidation, which consists of 53 species and 325 reaction steps (converted to 634 irreversible reactions). As in the previous section, we chose a global time step size of \SI{1e-6}{\second} and reported the average computational time for ten steps. Adjacent ODEs used similar initial conditions. In this case, RKC-GPU performed nearly eight times faster than RKCK-GPU in terms of computational time, suggesting more significant stiffness compared to Section~\ref{S:h2-co}. Consequently, we also compared the performance of RKC-GPU with the CPU-based implicit solver VODE. In both RKC and VODE, we selected a relative tolerance of \num{1e-6} and an absolute tolerance of \num{1e-10}.

\begin{figure}[tbp]
\begin{center}
\includegraphics[width=0.8\linewidth]{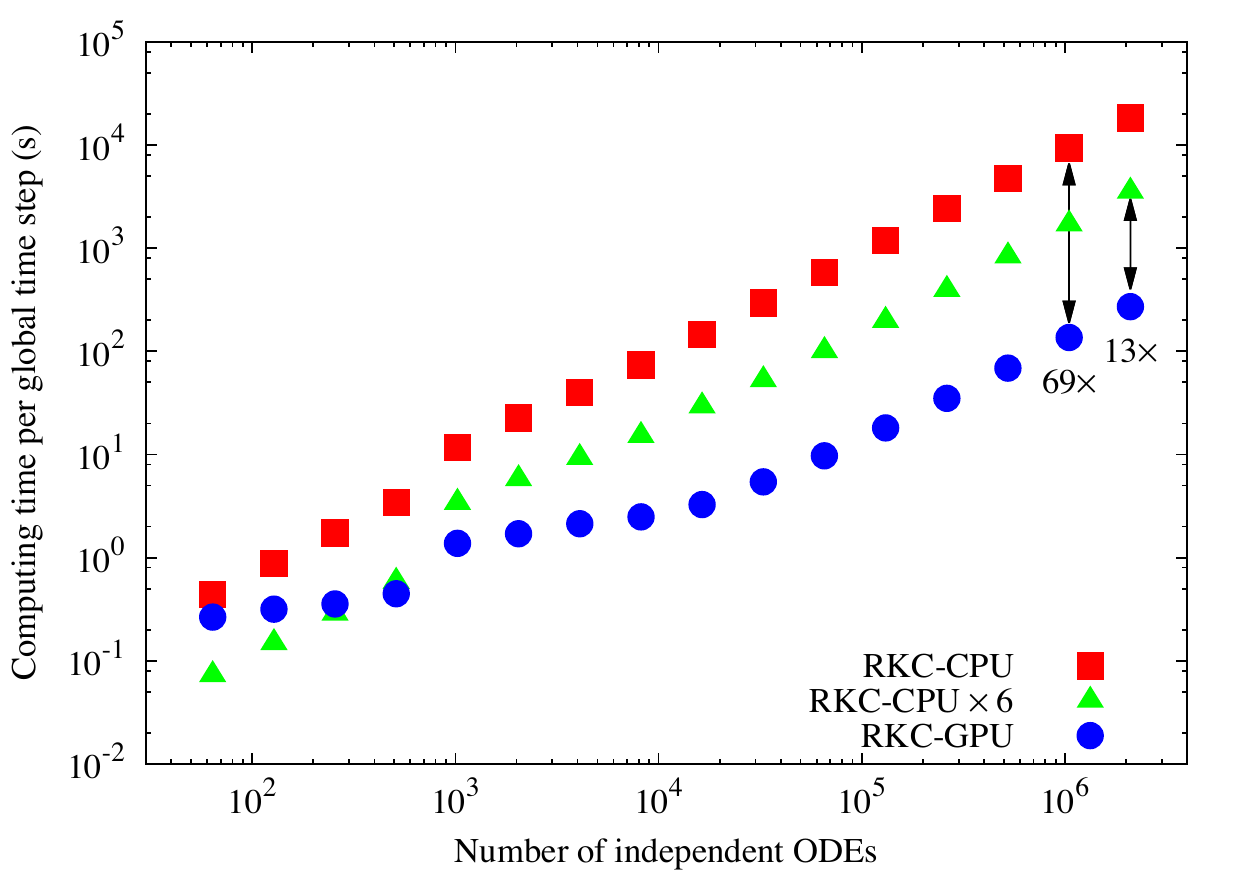}
\caption{Performance comparison of (single- and six-core) CPU and GPU integration of the moderately stiff methane mechanism using the stabilized explicit RKC method.}
\label{F:ch4-rkc}
\end{center}
\end{figure}

Figure~\ref{F:ch4-rkc} shows the performance comparison between the single- and six-core RKC-CPU and RKC-GPU for problem sizes ranging from 64 to \num{2097152}. As before, RKC-GPU performed better at larger problem sizes. Similar to the  hydrogen\slash carbon monoxide mechanism results, RKC-GPU outperformed RKC-CPU using a single CPU core for all ODE numbers considered here and faster than the six-core version for problem sizes of 512 and larger. At larger problem sizes, RKC-GPU compared slightly more favorably than in the previous section, performing up to 69$\times$ and 13$\times$ faster than RKC-CPU on one and six CPU cores, respectively. The jump in computational time between 512 and \num{1024} ODEs corresponded to the addition of initial conditions with greater stiffness.

\begin{figure}[tbp]
\begin{center}
\includegraphics[width=0.8\linewidth]{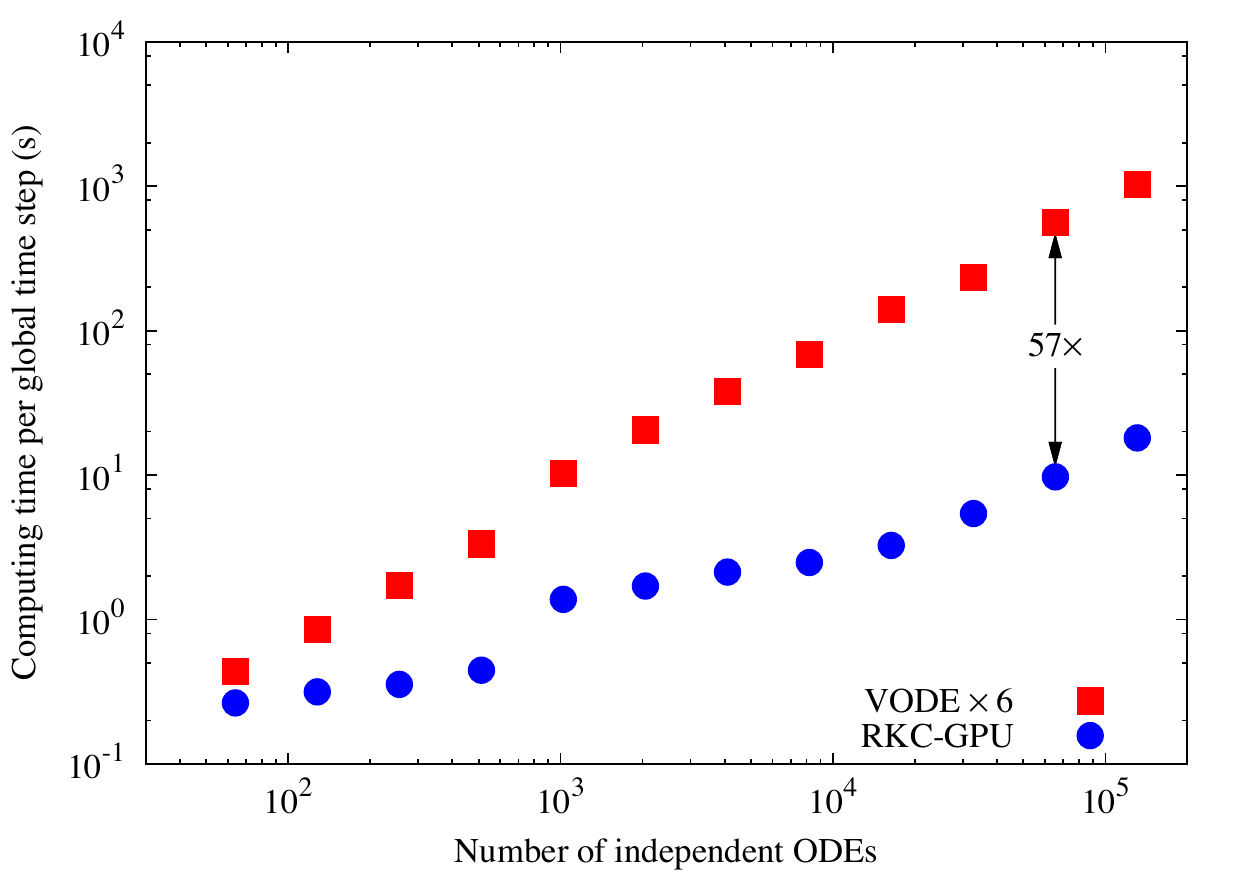}
\caption{Performance comparison between VODE running on six CPU cores and RKC-GPU with the moderately stiff methane mechanism over a wide range of problem sizes (i.e., number of ODEs).}
\label{F:ch4-rkc-vode}
\end{center}
\end{figure}

Since this problem exhibited greater stiffness compared to the previous case, we also studied the performance of VODE on the CPU compared against RKC-GPU. Figure~\ref{F:ch4-rkc-vode} shows the computational time for VODE on six CPU cores and RKC-GPU for problem sizes ranging from 64 to \num{1048576}. At all numbers of ODEs considered, RKC-GPU performed faster than VODE, demonstrating a speedup of up to 57$\times$. Though it is not shown here, we also note that RKC-CPU outperformed VODE---both on six CPU cores---by a factor of three to six.

\begin{figure}[tbp]
\begin{center}
\includegraphics[width=0.8\linewidth]{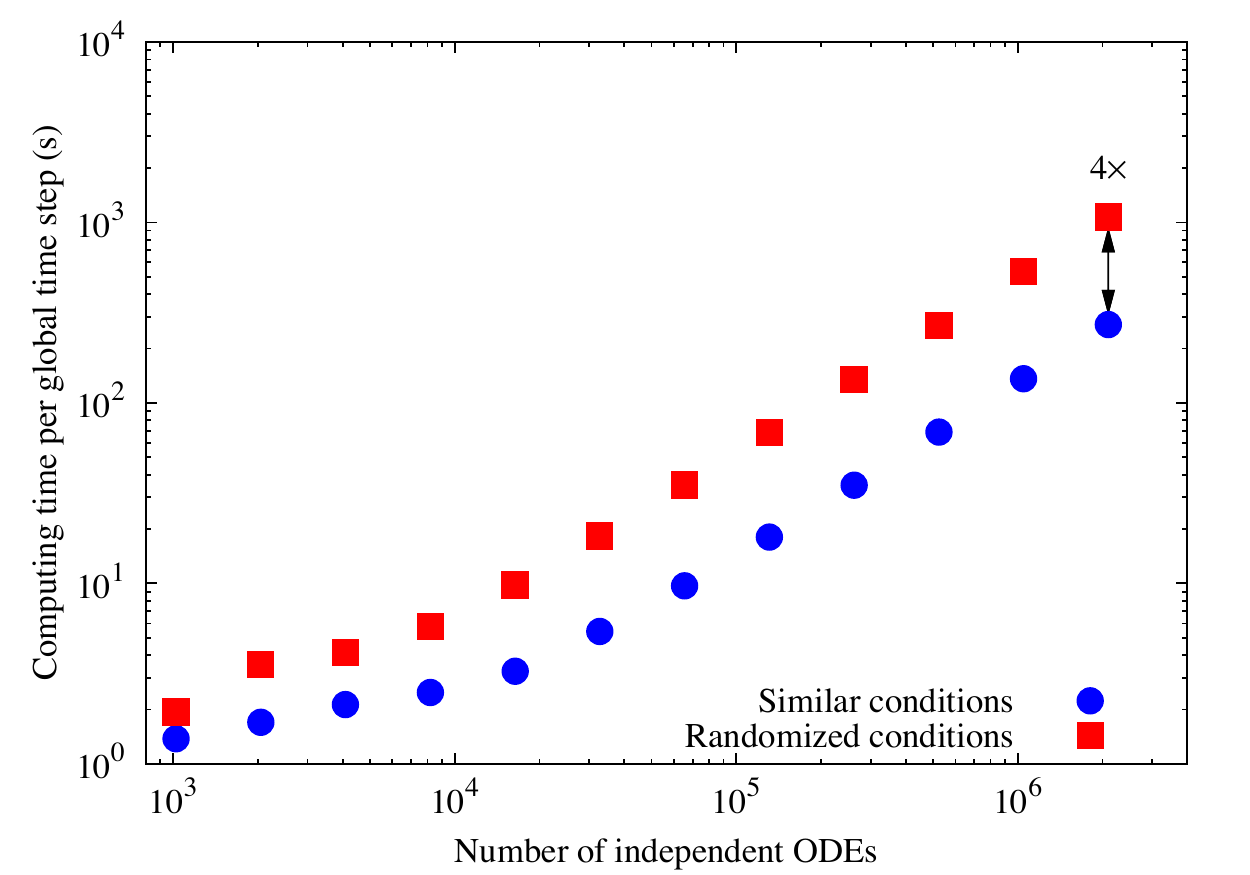}
\caption{Performance comparison of RKC-GPU integration of the methane mechanism where neighboring threads have similar and randomized initial conditions.}
\label{F:ch4-rkc-random}
\end{center}
\end{figure}

\begin{figure}[tbp]
\begin{center}
\includegraphics[width=0.8\linewidth]{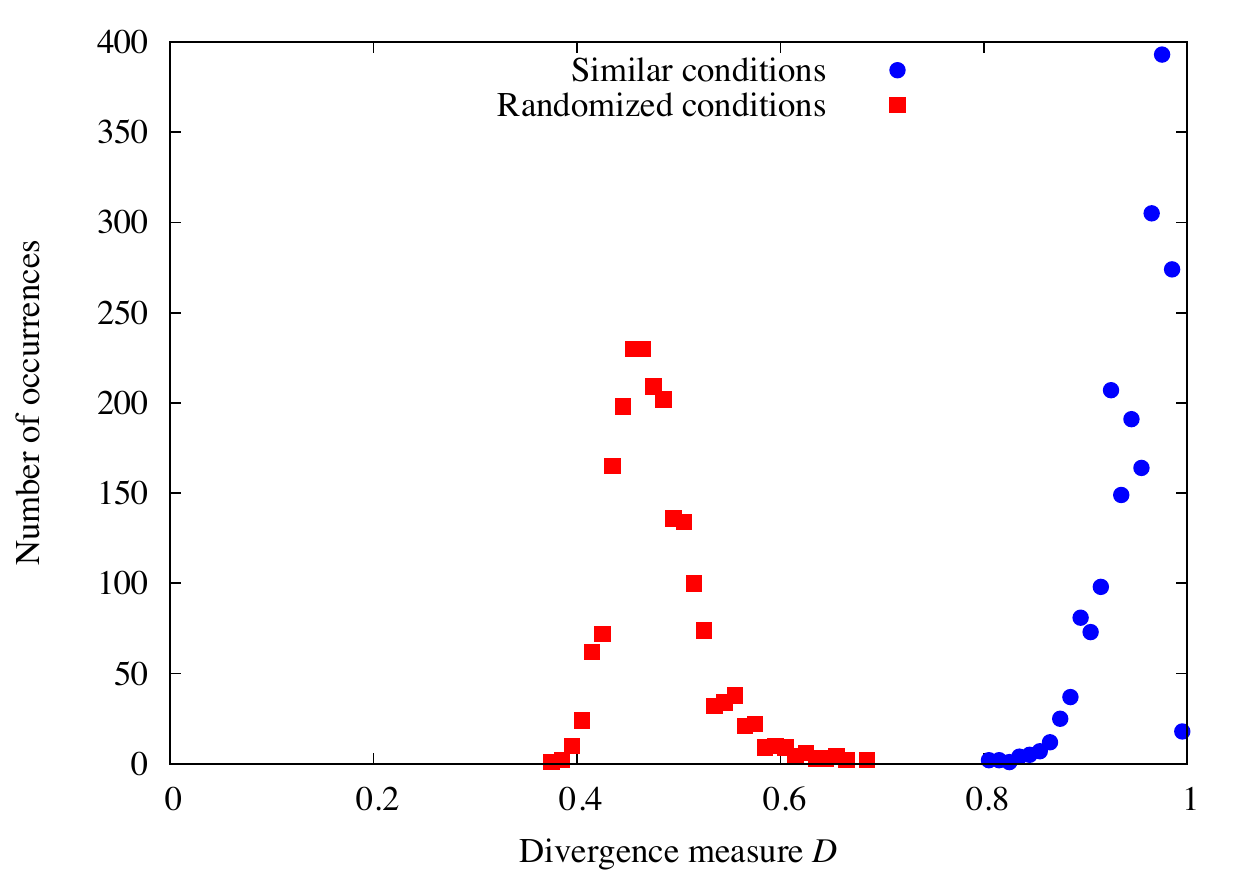}
\caption{Warp thread divergence comparison of RKC-GPU for methane kinetics for similar and random initial conditions, where the number of occurrences of the divergence measure \emph{D} is plotted for \num{2048} thread warps.}
\label{F:ch4-diverge}
\end{center}
\end{figure}

Figure~\ref{F:ch4-rkc-random} shows the performance of RKC-GPU for methane kinetics when the initial conditions in neighboring threads were similar and randomized. The behavior demonstrated here was similar to that in the previous section, with increasing disparity in performance for larger numbers of ODEs. In this case, with randomly selected initial conditions RKC-GPU performed up to nearly four times slower than when threads contained similar initial conditions. This drop in performance can also be seen in the distribution of \emph{D} for \num{65536} ODEs (\num{2048} warps) in Fig.~\ref{F:ch4-diverge}. The divergence, as measured by \emph{D}, showed similar behavior to that of hydrogen\slash carbon monoxide in Fig.~\ref{F:h2-rkc-diverge}. Here, for similar conditions, \emph{D} is clustered near one, and for randomized initial conditions normally distributed around 0.45---slightly lower than with the hydrogen\slash carbon monoxide mechanism. This likely explains the slightly greater drop in performance for randomly ordered initial conditions, compared to the previous section.

\subsection{Ethylene kinetics}
\label{S:ethylene}

Finally, we studied the performance of RKC-GPU in a case where stiffness is more severe: ethylene oxidation using the USC Mech version II mechanism~\cite{Wang:2007}, which consists of 111 species and 784 reactions (converted to \num{1566} irreversible reactions). In both RKC and VODE, we selected a relative tolerance of \num{1e-6} and an absolute tolerance of \num{1e-10}. Adjacent ODEs used similar initial conditions.

\begin{figure}[tbp]
\begin{center}
\includegraphics[width=0.8\linewidth]{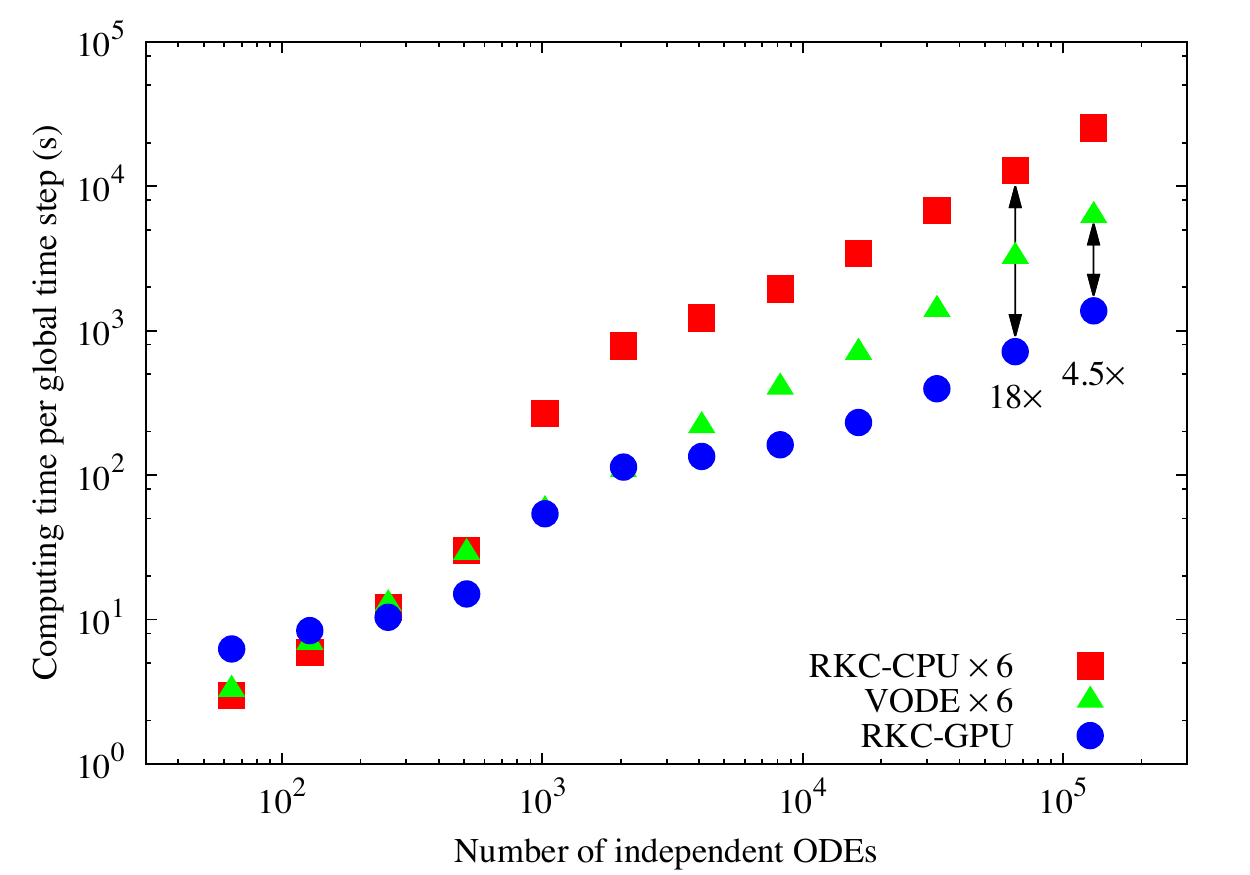}
\caption{Average computational time required for a \SI{1e-6}{\second} global time step using RKC-CPU, RKC-GPU, and VODE with the ethylene oxidation mechanism, for a wide range of ODE numbers. Both RKC-CPU and VODE were performed on six CPU cores.}
\label{F:c2h4-rkc-vode}
\end{center}
\end{figure}

Figure~\ref{F:c2h4-rkc-vode} shows the computational time for RKC-CPU, RKC-GPU, and VODE for numbers of ODEs ranging from 64 to \num{131072}. As before, we chose a global time step size of \SI{1e-6}{\second} and reported the average computational time for ten steps. Both CPU-based algorithms were executed on six CPU cores. Here, we omit the single-core RKC-CPU results; the performance ratio between the single- and six-core version showed similar scaling (4--6$\times$) to that shown in the previous sections. At problem sizes smaller than 256 ODEs, both RKC-CPU and VODE performed faster than RKC-GPU. RKC-GPU and VODE showed nearly indistinguishable performance for \num{1024} and \num{2048} ODEs. For numbers of ODEs greater than \num{8192}, RKC-GPU performed 12--18$\times$ faster than RKC-CPU and 2.5--4.5$\times$ faster than VODE.

\begin{figure}[tbp]
\begin{center}
\includegraphics[width=0.8\linewidth]{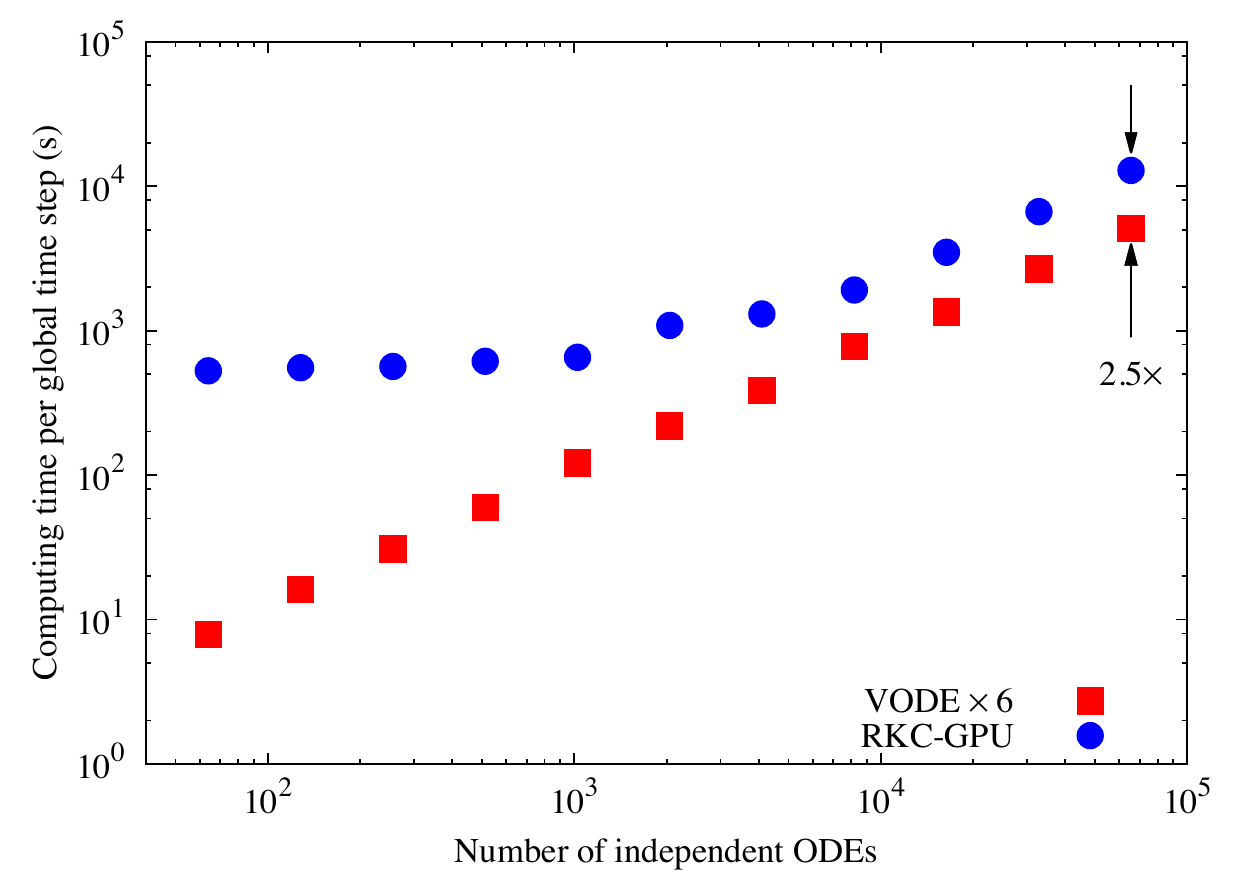}
\caption{Performance comparison between VODE running on six CPU cores and RKC-GPU with the ethylene mechanism over a wide range of problem sizes for a global time step size of \SI{1e-4}{\second}.}
\label{F:c2h4-1e-4}
\end{center}
\end{figure}

Next, we increased the global time step size to \SI{1e-4}{\second} to further increase the severity of stiffness. Figure~\ref{F:c2h4-1e-4} shows the performance of RKC-GPU and six-core VODE for numbers of ODEs ranging from 64 to \num{65536}. For all problem sizes here, RKC-GPU is slower than VODE. At best, RKC-GPU ran 2.5$\times$ slower than VODE for \num{16384} ODEs.

\subsection{Discussion}
\label{S:discussion}

The results shown above demonstrate that GPUs may be used to significantly reduce the cost of incorporating detailed chemistry in reactive-flow simulations. When stiffness is low due to the chemistry or small time step sizes used---such as those used in high-speed flow or DNS studies---explicit algorithms such as RKCK offer significantly higher performance on CPUs than implicit methods. Implementing RKCK on the GPU compounds this performance benefit over an order of magnitude, performing up to factor of 25 faster than the equivalent six-core CPU version in this study. As such, GPU-accelerated explicit methods are an attractive choice for nonstiff problems.

However, many chemical kinetics problems exhibit stiffness and therefore implicit algorithms are typically chosen to integrate the chemistry terms. As shown here, though, stabilized explicit methods such as RKC offer another option when stiffness is moderate. Demonstrated with methane kinetics, the RKC-GPU solver performed up to nearly 60$\times$ faster than the implicit VODE solver on six CPU cores. In fact, even the CPU implementation of RKC outperformed VODE. Based on these results, we suggest that a GPU-accelerated stabilized explicit method like RKC should be used in place of the standard implicit solvers in reactive-flow simulations---when stiffness is moderate. Typically, high-fidelity simulations use mechanisms with less than around 100 species---the size of those used in this study---so applying the GPU-based RKC integrator could significantly reduce the cost of chemistry in such studies. In addition, it could allow the use of larger, more complex mechanisms.

In the presence of more severe stiffness, as with the ethylene oxidation mechanism here, the GPU-accelerated RKC still showed significant speedup over the CPU version. Unfortunately, the comparison between VODE (on six CPU cores) and RKC-GPU became less favorable, with the speedup dropping to a factor of 4.5 for a global time step size of \num{1e-6}. The performance of RKC-GPU compared to VODE dropped further when the global time step size was increased---due to the greater stiffness this induced. For example, a time step size of \SI{1e-4}{\second} may be used for engine simulations; in this case, RKC-GPU performed at best 2.5$\times$ slower than VODE (on six CPU cores). As Stone et al.~\cite{Stone:2013jf} demonstrated, porting VODE to the GPU may not yield much benefit over a multi-core CPU implementation. Therefore, for problems with severe stiffness, an integration algorithm appropriate for GPU acceleration needs to be developed.

In all cases shown here, the speedup of the GPU-based algorithm compared to the equivalent CPU-based algorithm improved with increasing numbers of ODEs; at the smallest numbers, the six-core CPU-based algorithms performed better. This trend agrees with that observed in previous efforts using various integration algorithms~\cite{Niemeyer:2011uw,Shi:2012cl,Stone:2013jf}. At smaller problem sizes, the overhead due to memory transfer between the GPU and controlling CPU dominates, while at larger problem sizes the time required for actual computation comprises most of the total wall-clock time.

Further, we observed a general trend of increasing RKC-GPU to RKC-CPU speedup with increasing mechanism size. For mechanisms with 13, 53, and 111 species, RKC-GPU performed up to 10$\times$, 13$\times$, and 18$\times$ faster, respectively, than the six-core RKC-CPU for a global time step size of \SI{1e-6}{\second}.

We also found that the the performance of both RKCK-GPU and RKC-GPU dropped by factors of up to 2.5 and 4.0, respectively, when adjacent threads (corresponding to spatial locations) used randomly shuffled---rather than similar---initial conditions. This was due to divergence from threads following different instruction pathways, since different conditions will result in varying inner time step sizes. RKC-GPU exhibited greater thread divergence due to additional sources from the spectral radius estimation and varying number of stages, and correspondingly with randomized initial conditions this method displayed a larger reduction in performance compared to RKCK-GPU relative to the respective CPU versions. In general, we consider the case of threads with similar initial conditions more realistic, since in reactive-flow simulations---particularly with structured grids---neighboring volumes\slash grid points will contain similar thermochemical states. However, a reduction in performance due to divergence could result in some cases, such as with unstructured grids, where neighboring locations may not be stored consecutively in memory.

\section{Conclusions}
\label{S:conclusions}

In the present work we demonstrated new strategies for accelerating reactive-flow simulations using graphics processing units (GPUs). Most approaches for such simulations rely on the operator-splitting technique, which separates the chemistry and transport terms in each time step for separate evaluation. This results in a large number of ordinary differential equations (ODEs) governing the evolution of the species mass fractions for each discretized spatial location (i.e., grid point or volume) that need to be solved each time step. Here, we demonstrated that explicit algorithms used to integrate the chemistry ODEs in parallel on GPUs can perform significantly faster than equivalent CPU versions. We employed the explicit fifth-order Runge--Kutta--Cash--Karp (RKCK) and second-order Runge--Kutta--Chebyshev (RKC) methods for nonstiff and moderately stiff kinetics, respectively.

We studied the performance of the RKCK algorithm using a nonstiff hydrogen mechanism with with 9 species and 38 irreversible reactions~\cite{Yetter:1991}, and the performance of the RKC algorithm using three mechanisms with increasing sizes and levels of stiffness: (1) hydrogen\slash carbon monoxide with 13 species and 54 irreversible reactions~\cite{Burke:2011fh}, (2) methane with 53 species and 634 irreversible reactions~\cite{Smith:2010}, and (3) ethylene with 111 species and \num{1566} irreversible reactions~\cite{Wang:2007}. By comparing the performance of the CPU and GPU versions of RKCK and RKC, as well as the CPU-based implicit VODE solver, over a wide range of problem sizes (i.e., number of chemistry ODEs), we drew the following conclusions:
\begin{itemize}
\item For cases without stiffness, the GPU-based RKCK outperformed the six-core CPU version by a factor of 25 at best.
\item For cases with moderate levels of stiffness, the GPU-based RKC performed faster than the six-core RKC-CPU by, at best, factors of 10 with a hydrogen\slash carbon monoxide mechanism, 13 with a methane mechanism, and 18 with an ethylene mechanism.
\item In the presence of moderate stiffness in the methane mechanism, RKC-GPU outperformed the implicit VODE solver---on six CPU cores---by a maximum factor of 57.
\item For cases with moderate stiffness, even the CPU-based RKC outperformed VODE.
\item With increased stiffness in the case of the ethylene mechanism, RKC-GPU performed only 4.5$\times$ faster at best than VODE on six CPU cores.
\item When stiffness became more severe due to a larger time step size used with the ethylene mechanism, RKC-GPU became less efficient than six-core VODE, performing at best 2.5$\times$ slower.
\item At small problem sizes (less than 512 ODEs), the six-core RKC-CPU was more efficient, but RKC-GPU outperformed the serial (single-core) CPU version in all cases considered here.
\item Due to thread divergence, the performance of the GPU solvers degraded with randomized (and therefore different) initial conditions in adjacent memory locations, by up to a factor of four slower compared to using similar initial conditions.
\end{itemize}

Finally, we note that while we used a second-order accurate RKC algorithm here, higher order RKC methods exist. For example, Abdulle~\cite{Abdulle:2002ws} developed a fourth-order RKC with similar traits to the current method. Our future work will involve implementing these higher order algorithms where such accuracy is needed, as well as developing a GPU-based stiff integrator that can handle severe stiffness.

\section*{Acknowledgements}

This work was supported by the National Science Foundation under grant number 0932559, the US Department of Defense through the National Defense Science and Engineering Graduate Fellowship program, the National Science Foundation Graduate Research Fellowship under grant number DGE-0951783, and the Combustion Energy Frontier Research Center---an Energy Frontier Research Center funded by the US Department of Energy, Office of Science, Office of Basic Energy Sciences under award number DE-SC0001198.


\appendix

\section{Irreversible reaction mechanism converter}
\label{A:irrev}

In order to avoid the conditional statements associated with the equilibrium constants required to calculate reverse reaction rate coefficients during a simulation, we can calculate reverse Arrhenius parameters ($A_r$, $\beta_r$, $T_{a r} = E_r / \mathcal{R}$) a priori. Our procedure is similar to that of Tur\'{a}nyi~\cite{Turanyi:2003}. Evaluating the reverse rate coefficients via the forward rates and equilibrium constants at three temperatures ($T_1$, $T_2$, $T_3$) results in three equations to solve for the three unknown reverse Arrhenius parameters:
\begin{align}
	k_{r1} &= A_r \cdot \exp \left( \beta_r \cdot \ln T_1 - T_{a r} / T_1 \right) \\
	k_{r2} &= A_r \cdot \exp \left( \beta_r \cdot \ln T_2 - T_{a r} / T_2 \right) \\
	k_{r3} &= A_r \cdot \exp \left( \beta_r \cdot \ln T_3 - T_{a r} / T_3 \right)
\end{align}
Let $ x_1 = \ln T_1 $, $ x_2 = \ln T_2 $, and $ x_3 = \ln T_3 $. Then, dividing by \emph{A} and taking the natural logarithm of each side,
\begin{align}
	\ln \left( \frac{k_{r1}}{A_r} \right) &= \ln k_{r1} - \ln{A_r} \nonumber \\
							&= \beta_r x_1 - \frac{T_{a r}}{T_1} \label{e:rev_log1} \\
	\ln \left( \frac{k_{r2}}{A_r} \right) &= \ln k_{r2} - \ln{A_r} \nonumber \\
							&= \beta_r x_2 - \frac{T_{a r}}{T_2} \label{e:rev_log2} \\
	\ln \left( \frac{k_{r3}}{A_r} \right) &= \ln k_{r3} - \ln{A_r} \nonumber \\
							&= \beta_r x_3 - \frac{T_{a r}}{T_3} \label{e:rev_log3}
\end{align}
Subtracting Eq.~\eqref{e:rev_log2} from Eq.~\eqref{e:rev_log1} and Eq.~\eqref{e:rev_log3} from Eq.~\eqref{e:rev_log2} gives
\begin{align}
	\ln k_{r 1} - \ln k_{r2} &= \beta_r \left( x_1 - x_2 \right) - T_{a r} \left( \frac{1}{T_1} - \frac{1}{T_2} \right) \nonumber \\
					&= \beta_r \left( x_1 - x_2 \right) - T_{a r} \frac{T_2 - T_1}{T_1 T_2} \label{e:kr12} \\
	\ln k_{r 2} - \ln k_{r3} &= \beta_r \left( x_2 - x_3 \right) - T_{a r} \left( \frac{1}{T_2} - \frac{1}{T_3} \right) \nonumber \\
					&= \beta_r \left( x_2 - x_3 \right) - T_{a r} \frac{T_3 - T_2}{T_2 T_3} \label{e:kr23}
\end{align}
Then, solving Eq.~\eqref{e:kr23} for $T_{ar}$,
\begin{gather}
	\frac{T_2 T_3}{T_3 - T_2} \left( \ln k_{r2} - \ln k_{r3} \right) = \frac{T_2 T_3}{T_3 - T_2} \beta_r \left( x_2 - x_3 \right) - T_{a r} \nonumber \\
	T_{a r} = \frac{T_2 T_3}{T_3 - T_2} \left( \beta_r \left( x_3 - x_2 \right) - \ln k_{r2} + \ln k_{r3} \right) \label{e:Tar_0}
\end{gather}
Inserting Eq.~\eqref{e:Tar_0} into Eq.~\eqref{e:kr12} for $T_{ar}$, and letting $a_1 = \ln k_{r1}$, $a_2 = \ln k_{r2}$, and $a_3 = \ln k_{r3}$, leads to:
\begin{gather*}
	a_1 - a_2 = \beta_r (x_1 - x_2) - \frac{T_2 - T_1}{T_1 T_2} \frac{T_2 T_3}{T_3 - T_2} \left( \beta_r (x_2 - x_3) - a_2 + a_3 \right) \\
	a_1 - a_2 + \frac{T_2 - T_1}{T_1 T_2} \frac{T_2 T_3}{T_3 - T_2} (a_3 - a_2) = \beta_r \left( x_1 - x_2 - (x_2 - x_3) \frac{T_2 - T_1}{T_1 T_2} \frac{T_2 T_3}{T_3 - T_2} \right)
\end{gather*}
Then, solving for $\beta_r$,
\begin{gather}
	\beta_r = \frac{ a_1 T_1 \left( T_3 - T_2 \right) + a_2 T_2 \left( T_1 - T_3 \right) + a_3 T_3 \left( T_2 - T_1 \right)}{x_1 T_1 \left( T_3 - T_2 \right) + x_2 T_2 \left(T_1 - T_3 \right) + x_3 T_3 \left(T_2 - T_1\right)} \label{e:beta_r}
\end{gather}
$T_{ar}$ is then
\begin{equation}
T_{ar} = \frac{T_1 T_2 T_3 \left( a_1 \left(x_2 - x_3\right) + a_2 \left(x_3 - x_1\right) + a_3 \left(x_1 - x_2\right)\right)}{x_1 T_1 \left( T_3 - T_2 \right) + x_2 T_2 \left(T_1 - T_3 \right) + x_3 T_3 \left(T_2 - T_1\right)}
\end{equation}

To solve for $A_r$, substitute the expressions for $\beta_r$ and $T_{ar}$ into Eq.~\eqref{e:rev_log1},
\begin{align}
	a_1 &- \ln A_r = \beta_r x_1 - \frac{T_{ar}}{T_1} \nonumber \\
	\ln A_r &= a_1 - \beta_r x_1 + \frac{T_{ar}}{T_1} \nonumber \\
	&= \frac{ a_1 T_1 \left(x_2 T_2 - x_3 T_3\right) + a_2 T_2 \left(x_3 T_3 - x_1 T_1\right) + a_3 T_3 \left(x_1 T_1 - x_2 T_2\right) }{x_1 T_1 \left( T_3 - T_2 \right) + x_2 T_2 \left(T_1 - T_3 \right) + x_3 T_3 \left(T_2 - T_1\right)} \\
	A_r &= \exp \left( \ln A_r \right)
\end{align}

We wrote a Python tool implementing the above conversion for Chemkin-format reaction mechanisms. This software is freely available online~\cite{Niemeyer:2013im}.

\end{document}